\documentclass[11pt, oneside]{article}   	
\usepackage{jheppub}

\usepackage{graphicx}				
\usepackage{amsmath}								
\usepackage{color}
\usepackage{mathtools}
\usepackage{amssymb}

\usepackage{hyperref}

\hypersetup{
    colorlinks,%
    citecolor=blue,%
    filecolor=black,%
    linkcolor=blue,%
    urlcolor=blue
}

\usepackage{overpic}
\usepackage{array}
\usepackage{rotating}
\usepackage{datetime}

\newcommand{\bea}{\begin{eqnarray}}
\newcommand{\eea}{\end{eqnarray}}
\newcommand{\be}{\begin{equation}}
\newcommand{\ee}{\end{equation}}

\title{Carroll symmetry, dark energy and inflation}
                                          \author[a]{Jan de Boer,}
                                           \author[b] {Jelle Hartong,}
                                           \author[c,d]{Niels A. Obers,}
                                           \author[e]{Watse Sybesma,}
                                           \author[f]{and Stefan Vandoren}

 \affiliation[a]{Institute for Theoretical Physics and Delta Institute for Theoretical Physics, \\
University of Amsterdam, P.O.Box 94485 1090 GL
Amsterdam, The Netherlands}
\affiliation[b]{School of Mathematics and Maxwell Institute for Mathematical Sciences, University of Edinburgh, Peter Guthrie Tait Road, Edinburgh EH9 3FD, UK} 
\affiliation[c]{Nordita, KTH Royal Institute of Technology and Stockholm University, \\
Hannes Alfvéns väg 12, SE-106 91 Stockholm, Sweden} 
\affiliation[d]{The Niels Bohr Institute, Copenhagen University, \\  Blegdamsvej 17, DK-2100 Copenhagen {\O} , Denmark}
\affiliation[e]{Science Institute, University of Iceland, \\Dunhaga 3, 107 Reykjav\'{i}k, Iceland}
\affiliation[f]{Institute for Theoretical Physics, Utrecht University, \\ Princetonplein 5, 3584 CE Utrecht, The Netherlands}

\abstract{
Carroll symmetry arises from Poincar\'e symmetry upon taking the limit of vanishing speed of light. 
We determine the constraints on the energy-momentum tensor implied by Carroll symmetry and show that for energy-momentum tensors of perfect fluid form, these imply an equation of state ${\cal E}+P=0$ for energy density plus pressure. Therefore Carroll symmetry might be relevant for dark energy and inflation.
In the Carroll limit, the Hubble radius goes to zero and outside it recessional velocities are naturally large compared to the speed of light.
The de Sitter group of isometries, after the limit, becomes the conformal group in Euclidean flat space.
We also study the Carroll limit of chaotic inflation, and show that the scalar field is naturally driven to have an equation of state with $w=-1$. Finally we show that the freeze-out of scalar perturbations in the two point function at horizon crossing is a consequence of Carroll symmetry.

To make the paper self-contained, we include a brief pedagogical review of Carroll symmetry, Carroll particles and Carroll field theories that contains some new material as well.
In particular we show, using an expansion around speed of light going to zero, that for scalar and Maxwell type theories one can take two different Carroll limits at the level of the action. In the Maxwell case these correspond to the electric and magnetic limit. For point particles we show that there are two types of Carroll particles: those that cannot move in space and particles that cannot stand still.

}
\begin{document}
                                           \maketitle

\section{Introduction}
In cosmology, the equation of state of a perfect fluid determines the evolution of the universe. The parameter $w$ relating the pressure and energy density, $P=w{\cal E}$, of the fluid, fixes the scale factor $a(t)$ in the Friedmann-Lema\^{i}tre-Robertson-Walker (FLRW) metric via the Einstein equations. For instance, $w=1/3$ and $w=0$ correspond to ultra-relativistic matter and non-relativistic matter respectively. Generically we have $a(t)\sim t^{\frac{2}{3(1+w)}}$ for a spatially flat universe (which is what we will consider in this paper). The case $w=-1$ is special and produces an exponentially expanding universe $a(t)\sim e^{Ht}$ with $H$ the Hubble constant. In cosmological models based on a single scalar field of which the energy-momentum tensor takes the form of a perfect fluid, we have 
\begin{equation}\label{w_inflaton}
w=\frac{\frac{1}{2c^2}\dot \phi^2-V(\phi)}{\frac{1}{2c^2}\dot \phi^2+V(\phi)}\ ,
\end{equation}
where we suppressed the spatial derivatives $\partial_i\phi$ as they are treated as inhomogeneous fluctuations in perturbation theory. 
According to the inflationary paradigm, at early times the inflaton field moves slowly and one gets $w=-1$ since the time derivatives $\dot \phi$ are small compared to the potential energy $V(\phi)$.

There is a different way of saying the same thing, but it relates to the Carroll limit%
\footnote{The $c \rightarrow 0$ Carroll limit of the Poincar\'{e} group, being the opposite of the $c\rightarrow \infty $ Galilean (non-relativistic) limit, was first considered in \cite{Levy1965,Bacry:1968zf}.} that we discuss in this paper. Using the conjugate momentum fields $\pi_\phi=\frac{1}{c^2}\dot\phi$, we can rewrite \eqref{w_inflaton} as
\begin{equation}\label{w_inflaton2}
w=\frac{\frac{1}{2}c^2\pi_\phi^2-V(\phi)}{\frac{1}{2}c^2\pi_\phi^2+V(\phi)}=-1+\frac{\pi_\phi^2}{V}\,c^2+{\cal O}(c^4)\ .
\end{equation}
In the last equation we have taken $c\to 0$ and assumed that the potential is non-zero and stays finite in the $c\to 0$ limit, otherwise $w=+1$. The leading term in the $c\rightarrow 0$ expansion corresponds to a perfect fluid with $w=-1$ and the other terms are suppressed in terms of higher powers in $c$. It is important that in this limit the momentum $\pi_\phi$ stays finite instead of the quantity $\dot \phi$. Keeping $\pi_\phi$ finite when $c\to 0$ can only happen when $\dot\phi\to 0$, in other words, the field is slowly varying. This is precisely the regime in which inflation works, and we will work out the example of chaotic inflation in the last section of this paper. 


There is an intuitive argument as to why we might want to consider small values of the speed of light. According to the Hubble-Lema\^{i}tre law, the recessional velocity $v$ of an object with respect to an observer separated by proper distance $d$, is given by $v=Hd$. If the object is far outside the Hubble sphere defined by the Hubble radius $R_H=cH^{-1}$, its recessional velocity satisfies $v\gg c$. For a nice discussion and review of this argument, see e.g. \cite{Davis:2003ad}. 
Thus we are in an opposite limit of the Newtonian, or non-relativistic, regime. In cosmological terms, these super-Hubble scales define the regime in which Carrollian symmetry will arise, that is our claim. As we send the speed of light to zero, the Hubble radius goes to zero, so essentially the entire universe becomes super-Hubble and hence Carrollian. The Hubble radius defines the causal patch of an observer, and as the Hubble radius goes to zero, the theory becomes ultralocal, one of the main characteristics of Carrollian physics.

As we will show in this paper, Carroll symmetry applied to an energy-momentum tensor that takes the form of a perfect fluid implies ${\cal E}+P=0$, a necessary condition for dark energy. It is not a sufficient condition since both ${\cal E}$ and $P$ could  vanish in the Carroll limit. The case of a free relativistic scalar field without potential (so $w=+1$) is a concrete example of a situation in which both energy and pressure vanish when $c\to 0$. What we show in this paper is that Carroll fluids (energy-momentum tensors of the form of a perfect fluid) with nonzero energy density must satisfy $w=-1$ and therefore model dark energy. 
When $w$ is time dependent, as e.g. with scalar field inflation, it will be driven to $w=-1$ in the Carroll limit. Another consequence of Carroll symmetry is that scalar perturbations freeze out when they cross the horizon, as we demonstrate at the end of this paper.

The general philosophy we advocate is not only to understand the Carroll point itself, but to understand properties of the universe in an expansion in $c$ around the value $c=0$. We build up spacetime in terms of small Hubble cells, with radius $cH^{-1}$, 
the Hubble horizon, and as we expand away from $c\to 0$, we make the Hubble cells local patches in which observers live. For small values of $c$, the physics is ultralocal, and lightcones close up, which is the hallmark of Carroll spacetime geometry. As $c$ increases the Hubble radius grows, so that more and more degrees of freedom can enter the Hubble horizon and we build up the relativistic properties of the universe. 

A remark is in order on what we mean when we send the speed of light to zero, since only dimensionless parameters have physical meaning. Thus in practice we take limits in which
the ratio $c/v_{\rm c}$ goes to zero with $v_{\rm c}$ a characteristic velocity of the setup in question\footnote{This is not the same as an ultra-relativistic limit, where $c/v_{\rm c}\to 1$. At some places in the literature, Carroll limits are sometimes incorrectly called ultra-relativistic limits.}. When considering the dynamics of particles in the Carroll limit, $v_{\rm c}$ 
is obviously the velocity of the particle, while in e.g. Carrollian electrodynamics
it will be an appropriate combination of the size of the electric and magnetic fields. As explained above, in the context of cosmology we can consider the recessional velocity 
$v_c  = H d$ of an observer at some distance $d$
outside the de Sitter horizon. There might also be interesting situations in condensed matter models, in which $c$ is not the speed of light but instead a Fermi-velocity, or a sound velocity. Supersonic phenomena should then obey the constraints imposed by Carroll symmetry, we expect. 

The Carroll symmetry of the limiting point where $c=0$ may provide an organizing principle in the study of the perturbative expansion around it. 
For instance, this vantage point has recently proven to be beneficial in the study of the opposite limit, namely the non-relativistic limit $c\to\infty$. More concretely, a non-relativistic expansion of the relativistic Lie algebra reveals the underlying structure of the $1/c^2$ expansion in general relativity coupled to matter \cite{VandenBleeken:2017rij,Hansen:2020pqs,Hansen:2018ofj}.
Although we will not address exhaustively the $c^{2}$ expansion in the present work, 
we do adopt this approach to obtain useful insights into Carroll scalars and electrodynamics. Notably, we construct a Lagrangian that reproduces the magnetic sector of Carroll electrodynamics, which was previously unknown. More generally,  before turning to these field theory considerations, we briefly review as an aid to the reader  Carroll symmetry and Carroll particles. This part will also include some new results. In particular, we show that a zero energy particle always moves while a nonzero energy particle cannot move in space.

The Carroll limit and corresponding symmetry algebra were initially studied in \cite{Levy1965,Bacry:1968zf}. It turns out that this limit is also non-trivial from the following points of view, e.g., in \cite{Bergshoeff:2014jla} it was demonstrated that non-trivial dynamics for coupled Carroll particles can be realized and in \cite{Duval:2014uoa,Bagchi:2019xfx,Bagchi:2019clu,Banerjee:2020qjj} Carrollian field theories were studied. Models with tachyonic aspects respecting Carrollian symmetries were considered in, e.g., \cite{Gibbons:2002tv,Batlle:2017cfa,Barducci:2018thr} and will furthermore feature in this work.
Aspects of Carrollian gravity have received attention in e.g. \cite{Henneaux:1979vn,Teitelboim:1981fb,Nzotungicimpaye:2014wya,Hartong:2015xda,Bergshoeff:2017btm,Duval:2017els,Ciambelli:2018ojf,Morand:2018tke,Penna:2018gfx,Donnay:2019jiz,Gomis:2019nih,Ballesteros:2019mxi,Gomis:2020wxp,Grumiller:2020elf,Bergshoeff:2020xhv}. 
Another catalyst for research is due to the (conformal) Carroll group and its connection to the Bondi, van der Burg, Metzner, Sachs (BMS) group, which in turn plays a central role in understanding gravity in asymptotically flat space, see e.g. \cite{Bagchi:2013bga,Duval:2014uva,Duval:2014lpa,Bagchi:2015nca,Hartong:2015usd,Cardona:2016ytk,Bagchi:2016bcd,Ciambelli:2019lap,Bagchi:2019cay}.

As has been argued in this introduction, in the context of cosmology, the manifestation of Carroll symmetry in hydrodynamics can be important. Carroll fluids have previously been addressed in  \cite{deBoer:2017ing,Ciambelli:2018xat,Ciambelli:2018wre,Ciambelli:2020eba}. 

The outline of this article is as follows. 
In Section \ref{sec:carrollimit} we review some basics of Carroll transformations and representations of the Carroll algebra. 
We furthermore investigate the Carroll limit of de Sitter space.
In Section \ref{sec:CP} we treat Carroll particles and their realisation from an extended phase space approach. 
We delve into Carroll field theory in Section \ref{sec:CFT} and through an expansion in small $c$ we systematically obtain Carroll boost invariant field theoretical realisations of scalar fields and Maxwell fields. 
We notably present an action that yields the equations of motion for the magnetic Carroll section.
In Section \ref{sec:PF} we establish the general equation of state $\mathcal{E}+P=0$ for perfect Carroll invariant fluids.
The Carroll limit of the Friedmann equations is investigated in Section \ref{FE} and in Section \ref{Inflation} we analyze what happens to inflation in the Carroll limit.
Finally, in Section \ref{sec:PS} we study scalar perturbations during inflation and consider freeze out in the context of the Carroll limit.
We conclude with an outlook in Section \ref{sec:outlook}.



\subsubsection*{Note added}
As this work was nearing its completion the preprint \cite{Henneaux:2021yzg} appeared. In that work the authors consider a method to obtain inequivalent Carroll contractions of Poincar\'e invariant field theories at the level of the action. Our results of the Carroll `magnetic' and `electric' contractions of the scalar action and the Maxwell action, presented in Section \ref{sec:CFT}, overlap and agree with their results. However, rather than adopting a Hamiltonian perspective, we employ a complementary approach as we view these contractions as arising in the context of an expansion around $c\to0$.

\section{Carroll transformations and Carroll limits} \label{sec:carrollimit} 

\subsection{Carroll transformations and algebra}

It might seem counter-intuitive that the limit $c\to 0$ gives something non-trivial, but it is well-documented in the literature  (see e.g. Ref.~\cite{Levy1965,Bergshoeff:2014jla,Duval:2014uoa,Bagchi:2019xfx}) that such a limit can be taken on the Lorentz boosts to yield
\begin{equation}\label{CarrollBoost}
t'=t-\vec b \cdot \vec x\ ,\qquad {\vec x}^{\,'}=\vec x\ .
\end{equation}
This can easily be derived from the Lorentz boost generators $L_i\equiv \frac{1}{c}x^i\partial_t+ct\partial_i$. In the limit of $c\to 0$, and redefining the generators $C_i\equiv cL_i$, we find that $C_i\to x^i\partial_t$, which generate the Carroll boost transformations given in \eqref{CarrollBoost}.  We will take a detailed look at the Carroll limit in the next subsection.

An interesting consequence is that velocities transform under Carroll boosts as rescalings,
\begin{equation}\label{transfvel}
v'^{\,i}=\frac{{\rm d}x'^{i}}{{\rm d}t'}=\frac{v^i}{1-\vec b\cdot \vec v}\ .
\end{equation}
This implies one has to consider the cases of zero and nonzero velocities separately, as they are not related by Carroll boosts, contrary to the Lorentzian and Galilean cases. A useful quantity is the unit-norm velocity vector
\begin{equation}\label{unitvelocity}
n^i\equiv \frac{v^i}{|\vec v|}\ ,\qquad n'^{\,i}=\pm n^i\ ,
\end{equation}
which has the property that it stays finite in the zero velocity limit. It is boost invariant, up to a possible sign flip that can arise if the boost parameter is large enough, i.e. for $\vec b\cdot\vec v>0$. In fact, there is a discontinuity at $\vec b=\vec{v}/{v^2}$ above which the velocity changes direction under boosts. Actually there is a discontinuity for any $\vec b=\vec{v}/{v^2}+\vec b_\perp$ where $\vec b_\perp$ is orthogonal to $\vec v$.


The Carroll boosts together with the translations and spatial rotations form the Carroll algebra. The Hamiltonian in the Carroll algebra is a central charge. It commutes with all generators, and appears as a central charge in the commutator of translations and boosts:
\begin{equation}
[P_i, C_j]=\delta_{ij}H\ ,\qquad i=1,...,d\ ,
\end{equation}
with $H=\partial_t$ and $P_i=\partial_i$. Observe that the Hamiltonian is Carroll boost invariant since $\partial_t'=\partial_t$. This is to be contrasted with Lorentzian or Newtonian notions of energy, which transform under boosts. 


Because of translation symmetry in space and time, there will exist a conserved energy-momentum tensor. In the case of Lorentz symmetry, the energy-momentum tensor of a relativistic system is symmetric in the indices, and so the question arises: what is the equivalent for Carroll symmetry? The rotation symmetry of course still requires symmetry in spatial indices, but the Carroll boost will give something new.
 Writing the generators of the Lorentz and Carroll boosts as $L_i\equiv L^\mu_i\partial_\mu$ and $C_i\equiv C^\mu_i\partial_\mu$ respectively, we find from the Lorentz symmetry that the energy-momentum tensor is indeed symmetric
\begin{equation}\label{LorentzWard}
\partial_\mu\Big(T^\mu{}_\nu L^\nu_i\Big)=0\quad \rightarrow \quad \frac{1}{c}T^i{}_t+cT^t{}_i=0\ ,
\end{equation}
whereas from imposing Carrollian symmetry we get\footnote{This is an on-shell constraint, i.e. $T^i{}_t$ should only vanish upon using the equations of motion.} 
\begin{equation}\label{CarrollWard}
\partial_\mu\Big(T^\mu{}_\nu C^\nu_i\Big)=0\quad \rightarrow \quad T^i{}_t=0\ .
\end{equation}
Notice that \eqref{CarrollWard} follows from \eqref{LorentzWard} by taking $c\to 0$ while keeping $T^t{}_i$, finite. So the hallmark of Carroll symmetry is that $T^i{}_t=0$. In Section \ref{sec:PF}, we prove that for a perfect fluid, this implies ${\cal E}+P=0$.

It will be useful to have the Carroll-covariant transformation law on the energy-momentum tensor. Under a general coordinate transformation $x^\mu\rightarrow x'^\mu=x'^\mu(x)$ we have, $T'^{\,\mu}{}_\nu=\frac{\partial x'^{\mu}}{\partial x^\rho}\frac{\partial x^\lambda}{\partial x'^{\nu}}T^\rho{}_\lambda$. Under a Carroll boost, we find 
$T^{'i}{}_t=T^i{}_t$ and for the other components
\begin{equation}\label{eq:trafolaws}
T^{'t}{}_t=T^t{}_t-b_iT^i{}_t\ ,\qquad T^{'i}{}_j=T^i{}_j+b_jT^i{}_t\ ,
\qquad T^{'t}{}_i=T^t{}_i+b_iT^t{}_t-b_jT^j{}_i-b_ib_jT^j{}_t\ ,
\end{equation}
where we used the Carroll boost transformations  \eqref{CarrollBoost} and one can further simplify these expressions by setting $T^i{}_t=0$.

The fact that in Carroll spacetime $T^i{}_t=0$, implies that there is no energy flux possible and that the energy enclosed in any fixed volume is necessarily conserved. This is very strange, but as we shall see, it means that if the energy of a particle is non-zero, this particle cannot move (and there is only rest-energy), but if the particle can move, its energy must be zero. So it is possible to have momentum $T^t{}_i$, but no energy flux $T^i{}_t$. We illustrate this further in Figure 1.

\begin{figure}[h!]
\begin{center}
		\begin{overpic}[width=0.85\textwidth]{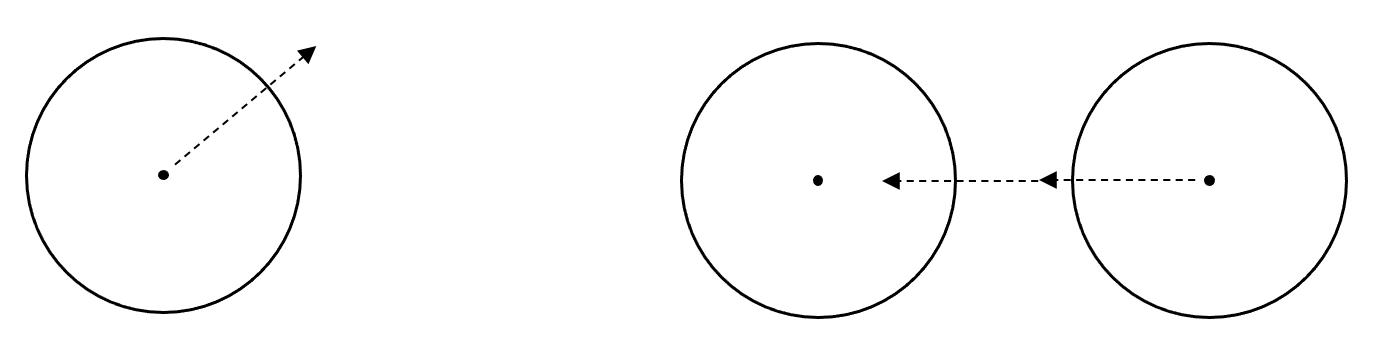}
			\put (3,3) {\footnotesize{$V$}}
			\put (10.5,10) {\footnotesize{$E$}}
			\put (50.5,3) {\footnotesize{$V_{1}$}}
			\put (79,3) {\footnotesize{$V_{2}$}}
			\put (58,10) {\footnotesize{$E_{1}$}}
			\put (87,10) {\footnotesize{$E_{2}$}}
		\end{overpic}
		
	\end{center}
\caption{Consequences of a zero energy flux. Left: consider a particle with energy $E$ enclosed by a volume $V$. If the particle can move, it could leave $V$ and energy inside $V$ is not conserved unless $E=0$. If the particle cannot move, then there can be a non-zero rest energy $E_0$ which stays inside $V$. Particle decay of a particle with non-zero rest energy can also not happen. Right: interactions are possible, but only between particles of zero and non-zero energy, or between particles with zero energy. In this figure, particle 2 is attracted to particle 1, but to be consistent with zero-energy flux through $V_2$, it must have vanishing energy $E_2=0$. When it enters $V_1$, the total energy inside $V_1$ is then still conserved. Particle 1 is all the time at rest with rest energy $E_1$. Under Carroll boosts, energies stay the same, but velocities are rescaled.}
\end{figure}
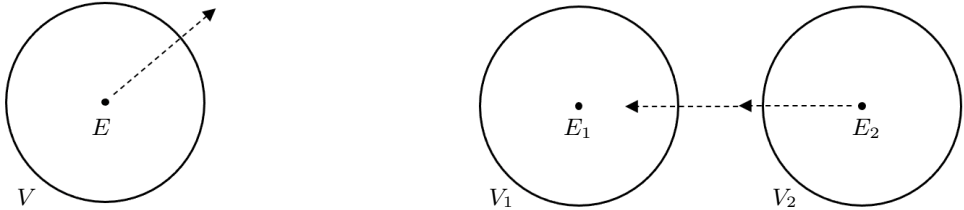

\subsection{Carroll transformations from Lorentz transformations}

In this subsection, we derive the Carroll transformations from the Lorentz transformations by taking the limit of vanishing speed of light. We denote the Lorentz boost transformation parameter by $\vec \beta$ with $0\leq |\vec {\beta}| <1$ and under a Lorentz boost we have

\begin{equation}
ct'=\gamma_\beta(ct-\vec \beta \cdot \vec x)\ ,\qquad {\vec x}^{\,'}_{\parallel}=\gamma_\beta({\vec x}_{\parallel} -\vec \beta ct)\ ,\qquad {\vec x}^{\,'}_\perp={\vec x}_\perp\ ,\qquad
\gamma_\beta=\frac{1}{\sqrt{1-\vec\beta^{\,2}}}\ ,
\end{equation}
with $\vec x={\vec x}_\parallel+{\vec x}_\perp$ the decomposition into components parallel and perpendicular to the boost parameter $\vec \beta$.
To get the Carroll transformations from this, we consider the scaling limit (following \cite{Duval:2014uoa}) 
\begin{equation}
\vec \beta \equiv c \,\vec b \ ,\qquad c\rightarrow \epsilon c\ ,\qquad \epsilon\rightarrow 0\ .
\end{equation}
The parameter $\vec b$ becomes the Carroll boost parameter, and its norm runs from zero to infinity. We keep it fixed in the limit $\epsilon \to 0$, and get 
\begin{equation}
t'=t-\vec b\cdot \vec x\ ,\qquad {\vec x}^{\,'}=\vec x\ .
\end{equation}
These are the Carroll boosts.
Under two consecutive Carroll boosts with parameters $\vec b_1$ and $\vec b_2$, we generate another one with boost parameter $\vec b_3=\vec b_1+\vec b_2$. Notice the difference with the non-relativistic Galilei limit which is obtained by setting $\vec\beta =c^{-1}\vec b$ and then sending $c\rightarrow \infty$. This gives the Galilei algebra as one can easily check.
In a Galilei universe, time is absolute, whereas in a Carroll universe, space is absolute. Since the Lorentz boost factor $\gamma_\beta\to 1$ in the Carroll limit, there is no Lorentz length contraction. There is no time dilation in a Carroll spacetime, but there is a time shift for events that are spatially separated. If a Carroll observer measures a time difference $\Delta t$ between two events separated by  $\Delta \vec{x}$, then a boosted Carroll observer measures the same distance, but a shifted time difference given by
\begin{equation}
    \Delta t'=\Delta t-\vec b\cdot \Delta \vec x\ .
\end{equation}
In other words, in a Carroll world time is relative, in contrast to the Galilean case, where space is relative. If a particle travels over a distance $\Delta \vec x$ in a time $\Delta t$, then after a boost with $\vec b\cdot \vec v>1$ with $\vec v=\Delta\vec x/\Delta t$, we get $\Delta t'<0$. What this all really means is that the coordinate time is not a good clock to describe the motion of a particle. Instead we will in the next section introduce a proper time, the affine parameter along the worldline of the particle. This proper time serves as an evolution parameter describing the motion of a particle.

Since we can have $\Delta t>0$ and $\Delta t'<0$, two Carroll observers do not necessarily agree on which event happened first. This sounds like a violation of causality, but it is not because for it to be a violation of causality physical information would have to be sent from one event to the other and since they are outside each other's Carroll light cone (as a result of the physical separation and the fact that light cones are lines at a fixed point in space) this would require a particle with a nonzero velocity. In a Carroll universe the latter is tantamount to a tachyonic particle. We will show further below that such particles exist and they in principle can lead to a violation of causality.


Spacetime distances become spatial distances in the $c\to 0 $ limit. The Minkowski metric degenerates as ${\rm d}s^2=-c^2{\rm d}t^2+{\rm d}{\vec x}^{\,2}\rightarrow {\rm d}{\vec x}^{\,2}$. Spatial distances are invariant under the Carroll group. Notice also that the Lorentz invariant $-c^2t^2+\vec {x}^{\,2}$ becomes the Carroll invariant $\vec {x}^{\,2}$ in the limit. Lightrays, for which this invariant is zero, in Carroll spacetime become $\vec x=0$ and $t$ arbitrary. Therefore, the coordinate $t$ parametrises the light cone. If $\vec x=0$, this means that light is not moving in space and the light cone has closed up.

Now we consider the contraction for the Lorentz transformations on energy and momentum
\begin{equation}
\vec{p}_\parallel{}^{'}=\gamma_\beta\left(\vec{p}_\parallel-\vec \beta\,\frac{E}{c}\right)\ ,\qquad
\vec{p}_\perp{}^{'}=\vec{p}_\perp \ ,\qquad
\frac{E '}{c}  = \gamma_\beta \left( \frac{E}{c} -\vec \beta\cdot \vec p\right)\ ,
\end{equation}
with $\vec p=\vec{p}_\parallel+\vec{p}_\perp$. In the Carroll limit we get
\begin{equation}\label{transfmomentum}
\vec{p}^{\,\,'}=\vec p -\vec b \,E \ ,\qquad E'=E\ .
\end{equation}
These are indeed the correct transformation rules in Carroll spacetime (as dictated by the Carroll algebra), and we have rederived that the Hamiltonian is Carroll boost invariant.

We can also look at the velocity of a particle, $\vec u\equiv {\rm d}\vec x/{\rm d}t$. Under Lorentz boosts, the parallel and perpendicular components transform as
\begin{equation}\label{transfu}
\vec u\,'_\parallel(t')=\frac{\vec u_\parallel - \vec \beta c}{\Big(1-\frac{\vec \beta \cdot\vec u}{c}\Big)}\ ,\qquad \vec u\,'_\perp(t')=\frac{\vec u_\perp(t)}{\gamma_\beta\Big(1-\frac{\vec \beta\cdot\vec u}{c}\Big)}\ .
\end{equation}
Taking the limit of \eqref{transfu} yields
\begin{equation}\label{transfvelocity}
\vec u^{\,'}(t')=\frac{\vec u(t)}{(1-\vec b\cdot \vec u)}\ ,
\end{equation}
which was also obtained directly in Carroll spacetime, see \eqref{transfvel}. As we mentioned there, the transformation law of velocity is quite strange. If the velocity was zero, it remains zero after a (finite) Carroll boost. These are the particles, considered e.g. in \cite{Bergshoeff:2014jla}. If it was non-zero, we can boost it to any other value, as large as we want. Hence we have two distinct classes, not related by Carroll boosts: particles with zero velocity and particles with non-zero velocity. We can even change the direction of the velocity vector under a large enough Carroll boost. That is by itself not surprising, but for small enough boost parameters with $\vec b\cdot \vec v < 1$, this will not happen.

Particles with non-zero velocity in the Carroll limit, satisfy $v>c\to 0$, and hence they are tachyonic. As discussed above, they can cause violation of causality. Clearly this is unphysical. The strict $c=0$ limit is unphysical. But this is not so different from the opposite limit, $c \to \infty$. The strict $c=\infty$ limit is an unphysical theory as well, as it causes action at a distance. It nevertheless is useful to study the symmetries and dynamics that emerge in this limit
. The proper way of thinking about both Carrollian and Galilei limits is to consider them as expansions in $c/v$ and $v/c$ respectively. We will discuss in more detail the dynamics of Carroll particles in the next section. 
\subsection{Representations of the Carroll algebra \label{subsec:rep}}

We will next discuss representations of the Carroll algebra in analogy with the massive and massless representations of the Poincar\'e algebra that are characterised by the eigenvalues of the momentum squared and the norm of the Pauli--Lubanski vector. Our results agree with those of  \cite{Duval:2014lpa}. 

The Carroll algebra consists of the generators $H$ (Hamiltonian), $P_i$ (spatial momenta), $C_i$ (Carroll boosts) and $J_{ij}=-J_{ji}$ (spatial rotations). The nonzero commutators are given by
\begin{eqnarray}
&&[P_i\,, C_j]=\delta_{ij}H\,,\\
&& [J_{ij}\,, P_k] = \delta_{ik}P_j-\delta_{jk}P_i\,,\\
&& [J_{ij}\,, C_k] = \delta_{ik}C_j-\delta_{jk}C_i\,,\\
&& [J_{ij}\,, J_{kl}] = \delta_{ik}J_{jl}-\delta_{jk}J_{il}+\delta_{jl}J_{ik}-\delta_{il}J_{jk}\,,
\end{eqnarray}
where $i,j,k,l$ run over $1,\ldots, d$. Representations are labelled by the eigenvalues of the central element $H$ and the quartic Casimir $M^2=\frac{1}{2}M_{ij}M_{ij}$ where 
\begin{equation}
M_{ij}=HJ_{ij}+C_i P_j-C_j P_i \,.
\end{equation}

We will momentarily specialize to $d=3$ and be quite explicit. Before doing so, we can 
describe the representations for general $d$ a bit more heuristically. There are basically
two cases, $H=0$ and $H\neq 0$. When $H=0$, $P$ and $C$ commute and so they can be simultaneously
diagonalized and states can be labeled by their eigenvalues. Rotations act on these eigenvalues 
and, possibly, simultaneously on an internal vector space. Once we extract an irreducible representation of the rotation algebra we can construct an irreducible representation of the Carroll algebra. In the absence of the internal vector space the zero energy irreducible representations can be labeled 
by $P_i P_i$, $P_i C_i$ and $C_i C_i$.
 
 For $H\neq 0$ the commutation relations of $P_i$ and $C_i$ are just like those of coordinates and
 momentum and can be represented by wave functions of $P_i$ or equivalently $C_i$. We can combine these wave functions with a representation $V$ of the rotation group, and then the rotation group acts simultaneously on this representation and by rotating $P_i$ and $C_i$. States in the 
 irreducible representation take the form $|P_i\rangle \otimes |\psi\rangle$ with $|\psi\rangle\in  V$.

We now describe the case of $d=3$ spatial dimensions more explicitly. 
For further details we refer to appendix \ref{app:irreps}. Here we summarise the main points. 

For $d=3$ it is useful to define $W_k$ and $S_k$ via
\begin{equation}
M_{ij}=\varepsilon_{ijk}W_k\,,\qquad J_{ij}=\varepsilon_{ijk}S_k\,,
\end{equation}
where $\varepsilon_{ijk}$ is the Levi-Civita symbol. In $d=3$ we can also define the operator $L=S_iP_i$ (which is closely related to the helicity operator). The Carroll algebra is a semi-direct sum of the Abelian ideal spanned by $\{H, P_i\}$ and the Euclidean algebra $\mathfrak{iso}(d)$ spanned by $\{J_{ij}, C_i\}$, so one can consider induced representations using the little group method. Our findings are:
\begin{itemize}
\item When $E\neq 0$ we can always go to a frame where $p_i=0$ by performing a Carroll boost. In this case the little group is $SO(3)$ and the eigenvalues of $W_i W_i$ are $E^2s(s+1)$ with $s=0,1/2,1,\ldots$. We can always go to a frame in which the states are of the form $|E\neq 0, \vec p=0, s, m\rangle$ where $m=-s,-s+1,\ldots,s$ is the eigenvalue of $S_3$, the spin along the $z$-axis.
\item When $E=0$ the momentum $p_i$ is Carroll boost invariant. Using a rotation we can without loss of generality set $\vec p=p\hat e_3$ where $\hat e_3$ is the unit vector along the $z$-axis. On such states $W_i=\varepsilon_{ijk} C_j P_k$ so that $W_3=0$. The little group is $ISO(2)$ and is generated by $W_1, W_2, L$. This case splits into two subcases. One for which $W_iW_i=0$ and one for which $W_iW_i>0$. We will always only consider the former. When $W_iW_i=0$ we can always go to a frame in which the states are of the form $|E=0, \vec p=p\hat e_3, \lambda\rangle$ where the helicity $\lambda$ is the eigenvalue of $S_3$.
\end{itemize}
Further below we will see examples of either of these two representations.

\subsection{Carroll limit of de Sitter space}\label{sec:carrollimitdS}

In gravity, spacetime is curved, so it is important to study the Carroll limit of Lorentzian metrics. For the purposes of this paper, we look at one particular example relevant for cosmology, namely de Sitter space consisting of pure dark energy. The exponentially expanding universe is described by the de Sitter metric, which in planar coordinates takes the form
\begin{equation}\label{dSmetricFLRW}
{\rm d}s^2=-c^2{\rm d}t^2+e^{2Ht}{\rm d}{\vec x}^{\,2}\ ,
\end{equation}
with $H$ the Hubble constant and with isometry group $SO(4,1)$ in four spacetime dimensions. The isometry group contains three rotations and three spatial translations, and furthermore the isometries corresponding to scale and special conformal transformations which infinitesimally take the form
\begin{eqnarray}
\delta t =-\frac{\alpha}{H}\ ,&\qquad & \delta \vec x=\alpha \vec x\ ,\\
\delta t =-\frac{\vec b\cdot\vec x}{H}\ ,&\qquad &\delta \vec x=(\vec b\cdot \vec x) \,\vec x+\frac{\vec b}{2}\Big(\frac{c^2}{H^2}e^{-2Ht}-{\vec x}^{\,2}\Big)\ .
\end{eqnarray}
All factors of $c$ are made explicit here, and so in the limit $c\to 0$, keeping $H$ fixed, the $b$-isometry becomes the Carroll isometry 
\begin{equation}\label{ConfCarroll}
\delta_C t =-\frac{\vec b\cdot\vec x}{H}\ ,\qquad \delta_C \vec x=(\vec b\cdot \vec x) \,\vec x-\frac{\vec b}{2}\,{\vec x}^{\,2}\ ,
\end{equation}
leaving the Carroll metric%
\footnote{More precisely, this is the spatial metric $h_{\mu \nu}$ of Carrollian geometry. The latter includes in addition the timelike vector (or inverse vielbein) $v^\mu$, satisfying $v^\mu h_{\mu \nu} =0$. See e.g. \cite{Hartong:2015xda,Bekaert:2015xua}.} 
${\rm d}s^2=e^{2Ht}{\rm d}{\vec x}^{\,2}$ invariant\footnote{This Carrollian spacetime is a homogeneous spacetime, referred to as the light cone in \cite{Figueroa-OFarrill:2018ilb}.}. In fact, this isometry is the infinitesimal special conformal transformation on $\mathbb{R}^3$, and the complete isometry group of this Carroll metric is the conformal group on $\mathbb{R}^3$. This hints towards a holographic description of (3+1)-dimensional de Sitter cosmology in terms of a three-dimensional Euclidean conformal field theory \cite{Strominger:2001pn,Strominger:2001gp}
(see also \cite{Maldacena:2011nz,Arkani-Hamed:2018kmz}), and  generalizes straightforwardly to other dimensions. We obtained it here in the limit $c\to 0$, but one finds the same conformal transformations at late times, $t\to \infty$. This suggests that late time de Sitter cosmology is governed by Euclidean conformal correlators, and this conformal symmetry can be understood from the Carroll symmetry that is present at super-Hubble scales. 

The metric in \eqref{dSmetricFLRW} is written in planar coordinates and they fit into the FLRW ansatz. One can also use conformal time, $\tau=-1/(aH)$, then we obtain the Carroll metric
\begin{equation}
    {\rm d}s^2=\frac{-c^2{\rm d}\tau^2+{\rm d}{\vec x}^2}{H^2\tau^2}\quad\stackrel{c\to 0}{\longrightarrow}\quad\frac{1}{H^2\tau^2}{\rm d}{\vec x}^2\ ,
\end{equation}
which is again the metric on three-dimensional Euclidean space, up to a conformal factor.

\section{Carroll particles}\label{sec:CP}

In this section we consider various aspects of Carroll particles.
Point particles can conveniently be described in an extended phase space system, which is useful
for Carroll particles (see e.g. also \cite{Bergshoeff:2014jla})  as generic Carroll systems do not seem to have a well-defined initial value problem
with respect to coordinate time. This follows from the fact that,
as discussed in the previous section,  Carroll transformations can change the direction of trajectories and even make them frozen in time. To have a well-defined initial value problem one can employ particles which use some internal clock time (cf. proper time) which can be used to define dynamics. 

\subsection{Extended phase space}

We first briefly recap how the extended phase space formulation works for ordinary particles. Consider the worldline action for a particle given by
\be
S_0=-\int {\rm d}s \,(E\cdot\dot{t} - \vec{p}\cdot\dot{\vec{x}})\ ,
\ee
which describes an extended phase space with $d+1$ coordinates and $d+1$ momenta ($d$ equals the number of spatial coordinates) and zero Hamiltonian. Notice that the combination $E\cdot\dot{t} - \vec{p}\cdot\dot{\vec{x}}$ is invariant under both Lorentz and Carroll boosts.

To reproduce some familiar systems from this we consider the class of actions 
\be
\label{psact}
S=-\int {\rm d}s \, (E\cdot\dot{t} - \vec{p}\cdot\dot{\vec{x}} - \lambda_\alpha O_\alpha +H)\ ,
\ee
where $O_\alpha$ are some constraints with Lagrange multipliers $\lambda_\alpha$ and $H$ is some Hamiltonian on the extended phase space. In order for this to make sense we need that $\{H,O_\alpha\}=c_{\alpha \beta} O_\beta$ so that time evolution preserves the constraint surface. We can have first class (commuting on the constraint surface) or second class constraints. In the first case the Lagrange multipliers will not be fixed as a reflection of the gauge symmetry which first class constraints typically  generate. In the second case the Lagrange multipliers will typically be fixed by the constraints and the equations of motion, and we can pass to the reduced phase space. 
One way to find the actual dynamics of the system, is by examining the resulting field equations. 

If the system has a set of symmetry generators on the extended phase space, the constraints and Hamiltonian must be compatible with these symmetries in order to preserve them. This can be studied canonically (i.e. through Poisson brackets of $O_\alpha$ and $H$ with the symmetry generators) but also through the action (i.e. verifying that it is invariant under these symmetries). Now, let us consider some cases:

\begin{enumerate}
\item[1.] $O_1=E-p^2/2m, H=0$. 
This includes a first class constraint. The field equations are 
\be
\dot{t}=\lambda_1,\quad \dot{E}=0,\quad \dot{\vec{p}}=0,\quad \dot{\vec{x}}=\frac{\lambda_1}{m} \vec p\ ,
\ee
and one sees that these still depend on $\lambda_1$. The action is $s$-reparametrization invariant, where one also needs to transform $\lambda_1$. We can
eliminate $\lambda_1$ to obtain ${\rm d} E/{\rm d} t = {\rm d} \vec{p}/{\rm d} t =0$ and $\vec{p}=m{\rm d} \vec{x}/{\rm d} t$ where everything 
depends on $t$ rather than $s$. Moreover $E=p^2/2m$. This describes a standard free massive non-relativistic particle. Eliminating the constraints in the action gives the usual form
\be
S=\int {\rm d}t\, \frac{p^2}{2m}=\frac{m}{2}\int {\rm d}t\,\dot{\vec{x}}^{\,2}\ ,
\ee
where the dot in the last equation now stands for the derivative with respect to $t$.
\item[2.] $O_1=E-p^2/2m, O_2=t-s; H=0$. 
We now have second class constraints and 
the field equations are 
\be
\dot{t}=\lambda_1,\quad \dot{E}=-\lambda_2,\quad \dot{\vec{p}}=0,\quad \dot{\vec{x}}=\frac{\lambda_1}{m}\, \vec{p}\ .
\ee
If we  combine these with the constraints we find $\lambda_2=0$ and $\lambda_1=1$, so the
Lagrange multipliers are completely fixed. Again, one finds a standard non-relativistic particle, since the extra constraint  simply relates $t$ to 
$s$ but does not have any essential impact. 
\item[3.] $O_1=E,H=p^2/2m$. This includes a first class constraint plus a non-trivial Hamiltonian. 
The field equations are 
\be
\dot{t}=\lambda_1,\quad \dot{E}=0,\quad \dot{p}=0,\quad \dot{\vec{x}}=\vec{p}/m\ ,
\ee
which is once more a standard particle. However, $t$ is no longer the usual time variable, as this role has been taken over
by $s$. In fact the dynamics in $t$ and $E$ completely decouples. The energy is not given by $E$ but by the Hamiltonian $H$. 
\item[4.] $ O_1=E-{\cal H}(p,x), H=0$. 
This is a another first class constraint and case 1 above is a special case of this. Rather than looking at the field equations, we integrate out $E$ and $\lambda_1$ which yields $\dot{t}=\lambda_1$ 
and $E={\cal H}(p)$. Plugging this into the action we obtain 
\be
S=\int {\rm d}t \left(\vec{p}\cdot\frac{{\rm d} \vec{x}}{{\rm d} t}-{\cal H}(p,x) \right) \ ,
\ee
which is just the Lagrangian in first order form. So such a first class constraint corresponds to standard dynamics in general.
\end{enumerate}

\subsection{Carroll symmetry and single Carroll particles}

We now apply the extended phase space technique to the dynamics of Carroll particles, starting with a single particle. The standard Carroll algebra has generators 
\be
E \ ,\quad Ex_i\ , \quad p_j \ ,\quad x_{[i} p_{j]}\ ,\quad i,j=1,...,d\ ,
\ee
for time translations, Carroll boosts, space translations and rotations respectively. It follows again that $E$ commutes with the other generators, and hence is a central charge.
One could imagine more complicated realizations of this algebra on a given phase space, but we will
restrict to this simple choice. 

To add a constraint consistent with Carroll symmetry, it must Poisson commute with these generators. Because of time
and space translational invariance (or Poisson commutability with $E$ and $\vec{p}$) the constraint can only depend on 
$E$ and $\vec{p}$, and rotational invariance restricts it to a function of $E$ and $p^2$. But this can only commute
with the generators $E\vec{x}$ if it does not depend on $p^2$. It follows  that $E-E_0$ is the only constraint that we can introduce. 
This produces the rather trivial case of a particle with energy $E_0$ and unconstrained momenta, which is also seen in the representation theory in section \ref{subsec:rep}. The extended phase space action based on this constraints reads
\begin{equation}
    S=-\int {\rm d}s \, (E\cdot\dot{t} - \vec{p}\cdot\dot{\vec{x}} - \lambda (E-E_0))\ ,
\end{equation}
giving rise to the equations of motion
\be
\dot t =\lambda\ ,\quad \dot E=0\ ,\quad \dot{\vec{x}}=0\ ,\quad \dot{\vec{p}}=0\ .
\ee
This corresponds to the Carroll particle at rest. It has zero velocity and constant energy $E_0$. This is to be expected since in the Carroll limit, the light-cone closes and only particles with zero velocity can survive the $c\to 0$ limit. 

There is, however, an interesting alternative choice corresponding 
to the introduction of  two constraints, $E=0$ and $p^2=p_0^2$.
This is consistent because the commutator of
the constraint $p^2-p_0^2$ with $E\vec{x}$ is proportional to the first constraint. This possibility was not included in the phase space analysis of \cite{Bergshoeff:2014jla}. It is easy to write down an action for this case. Using the extended phase space action 
\eqref{psact}  with constraints $O_1=E, O_2=p^2-p_0^2$ and $H=0$ for some constant $p_0$. One can subsequently eliminate the Lagrange multipliers to get $\lambda_1=\dot t$ and
\be
\lambda_2=-\frac{1}{2p_0}{{\sqrt{\dot{\vec x}^{\,2}}}}\ ,
\ee
such that the action becomes 
\begin{equation}\label{E0action}
S[\vec x]=p_0\int {\rm d}s\,{\sqrt {\dot{\vec{x}}\cdot \dot{\vec x}}}\ ,\qquad L(\vec x, \dot{\vec x})= p_0 {\sqrt {\dot{\vec{x}}\cdot \dot{\vec x}}}\ .
\end{equation}
Indeed, from the Lagrangian follows the momentum
\begin{equation}\label{eq:pandn}
\vec p=\frac{\partial L}{\partial \dot{\vec x}}=p_0\,\frac{\dot{\vec x}}{{\sqrt{\dot{\vec x}^{\,2}}}}=p_0\, \vec n\ ,
\end{equation}
which satisfies the contraint
\begin{equation}
O_2 \equiv \vec p^{\,\,2}-p_0^2=0\ ,
\end{equation}
as it should. The equation of motion says that $\vec p$ and therefore $\vec n$ is constant in time. $E$ and $t$ decouple from the dynamics and there is still reparametrization symmetry on the worldline. One can use this to fix a gauge $|\dot{\vec{x}}|=p_0$. Then the particle trajectories are of the form $\vec{x}(s)=\vec{p}_0s+\vec{x}_0$ for some vector $\vec{p}_0$ with $|\vec{p}_0|=p_0$. The solution for the momentum then becomes $\vec p=\vec{p}_0$, and we recall that the energy is zero. 

We thus confirm the properties concluded  in the previous section  as a consequence of having a zero energy flux: If particles can move, they must have zero energy, and if they have non-zero energy, they cannot move and there is only rest energy.
Having established this, we now turn to the question of how these
particles can arise from a relativistic theory by taking 
the limit $c \to 0$. 


\subsection{Carroll particles from relativistic particles}\label{tachyonCarroll}

Let us consider a free relativistic particle with energy $E$, mass $m$ and momentum $\vec p$. We denote the velocity of the particle by  $\vec u={\rm d}\vec x/{\rm d}t$ and $u^2=\vec{u}\cdot\vec{u}$. The relations between them are given by
\begin{equation}\label{p&E}
\vec p=\frac{m\vec u}{{\sqrt{1-\frac{u^2}{c^2}}}}\ ,\qquad E=\frac{mc^2}{{\sqrt{1-\frac{u^2}{c^2}}}}\ ,\qquad E^2=\vec{p}^{\,2}c^2+m^2c^4\ ,
\end{equation}
and hold in any Lorentz frame.

We now encounter a subtlety when taking the Carroll limit. The gamma-factor associated with the particle with velocity $\vec u$ does not seem to have a well-defined limit
\begin{equation}
\gamma_u=\frac{1}
{{\sqrt{1-\frac{u^2}{c^2}}}}\quad\stackrel{c\to 0}{\rightarrow} \quad ?
\end{equation}
Notice that this is a different object compared with the gamma-factor in the Lorentz transformation that contains the boost parameter $\vec\beta$, which always goes to unity in the Carroll limit: $\gamma_\beta\rightarrow 1$. Here we are not contracting the boost parameter, instead we are searching for a consistent limit of physical quantities under a Carroll contraction. 

If we keep $u$ fixed and non-zero, this expression becomes imaginary in the Carroll limit. It seems there are two options: either we require that the Carroll velocity identically vanishes, $u=0$, and then $\gamma_u\rightarrow 1$, or we allow for imaginary values 
\begin{equation}\label{branchgamma}
\gamma_u\rightarrow \mp\,i\,\frac{c}{u}\ ,\qquad u\equiv |\vec{u}|\ ,
\end{equation}
where we allow for the two branches of the square root.
This expression for $\gamma_u$ vanishes in the strict Carroll limit, but this is the leading term in the small $c$ expansion.

The case of zero velocity, $\vec u=0$ corresponds to the Carroll particle at rest. The limit $c\to 0$ is rather trivial, but we take the limit such that there is still rest energy $E_0$. To achieve this we keep the combination $mc^2$ fixed in the Carroll limit. The energy and momenta for these particles then are
\begin{equation}
    E=E_0=mc^2\ ,\qquad \vec p=0\ .
\end{equation}

We now consider the more interesting case in which $\gamma_u=\mp i\,\frac{c}{u}$. Substituting this into the expression for the momentum, we find
\begin{equation}
\vec p=\mp(imc)\frac{\vec u}{u}=\mp (imc)\vec n\ .
\end{equation}
This can only make sense and yield real and non-zero values of the momentum if we keep $mc$ fixed and the original relativistic particle was a tachyon, so
\begin{equation}
p_0\equiv -imc\ ,
\end{equation}
for some real $p_0$ and therefore $m^2<0$. That tachyons enter into the picture is not so surprising since \eqref{branchgamma} also follows from $u^2>c^2$. Notice furthermore that the scaling law on the mass $m$ is different as compared to the case of the relativistic massive particle with zero velocity, where the mass $m$ scales like $\epsilon^{-2}$ in the Carroll limit. The momentum of these tachyon-like particles satisfies the constraint $p^2=p_0^2$ since $\vec n$ is of unit norm. Furthermore, we find that the Carroll particle has zero energy because of the cancellation $\vec p^{\,2}c^2+m^2c^4=\vec p^{\,2}c^2-p_0^2c^2=0$ (thanks to the tachyon!), and hence the energy and momenta for these particles are
\begin{equation}\label{eq:pandn2}
E=0\ ,\qquad \vec p=\pm p_0\vec n\ .
\end{equation}
Notice the $\pm$ sign in the momentum, which is a consequence of the fact that the momentum as a function of the velocity is no longer single-valued (as follows from \eqref{branchgamma}). The velocity $u$ is undetermined and arbitrary but non-vanishing. These Carroll particles, for $p_0\neq 0$, cannot be put to rest by a boost, they always move. 

It is interesting to contrast equation \eqref{eq:pandn2} with \eqref{eq:pandn}. The latter does not have a sign ambiguity. To go from \eqref{eq:pandn} to \eqref{eq:pandn2} we need to divide the numerator and the denominator of \eqref{eq:pandn} by $\dot t$. However $\dot t$ can have either sign which is the origin of the sign ambiguity of $\vec p$ when expressed in terms of the velocity $\vec u=\dot{\vec x}/\dot t $. This illustrates what was stated previously, namely that the more convenient evolution parameter is the proper time $s$ as opposed to coordinate time $t$.

 A similar representation with zero energy can be found from the Carroll limit of massless particles. These have dispersion relation $E=pc$, and for fixed $p$, we have $E\to 0$ in the Carroll limit.

\subsection{Many-particle Carroll systems}

It is relatively easy to generalize the extended phase space approach to many particles which we label by $a=1,...,N$. The action on extended phase space is now
\begin{equation}
    S=-\int ds\left(E_a\dot t_a-\vec p_a\cdot\dot{\vec x}_a-\lambda_\alpha O_\alpha+H\right)\,,
\end{equation}
where we sum over $a$ (the number of particles) and $\alpha$ (the number of constraints). To get a standard classical mechanical system of many particles, we can either choose a set of first class constraints
$E_a-{\cal H}(\vec{x}_a,\vec{p}_a)=0$ for all $a$ with ${\cal H}$ the total Hamiltonian (that we assume to be time independent) and $H=0$, or choose first class constraints $E_a=0$ and add an explicit Hamiltonian $H$ just as we did for the single particle case. In either case, using the gauge symmetries, this leads to standard actions for multiple particles with a single time variable $t=\sum_{a=1}^Nt_a$.

We now apply this to a system of many Carroll particles, as was also considered in Ref.~\cite{Bergshoeff:2014jla}. 
In the case of $N$ particles, and can realise the generators of the Carroll algebra as
\be
\sum_a E_a \,, \quad \sum_a E_a\vec{x}_a\,, \quad \sum_a \vec{p}_a\,, \quad \sum_a x^{[k}_a p^{l]}_a\,,\quad a=1,...,N\ .
\ee
Apart from rotations, the following set of building blocks commute with the Carroll algebra:
\be \label{inv1}
 E_a\,, \quad \vec{x}_a - \vec{x}_b\,, \quad E_a \vec{p}_b - E_b \vec{p}_a\,,
\ee
but also the combinations
\be \label{inv2}
E_a t_{bc} - \vec{p}_a  \cdot \vec{x}_{bc}\ ,
\ee
with $t_{ab}=t_a-t_b$ and $\vec{x}_{ab}=\vec{x}_a-\vec{x}_b$. The building blocks \eqref{inv1} and \eqref{inv2} form a closed algebra under the Poisson bracket\footnote{The basic Poisson brackets on extended phase space are $\{t\,,E\}=-1$ and $\{x^i\,,p_j\}=\delta^i_j$.} and are thus first class.

To construct an interacting Carroll system, we need to find a set of commuting constraints, or as shown above, a set of commuting 
constraints plus a Hamiltonian. In fact, we only need that the variation of the Hamiltonian and the constraints is 
proportional to the constraints, as in that case we can make the theory invariant by letting the Lagrange multipliers transform. 


For Carroll symmetry this means that we can take first class constraints of the form $E_a-H$ where $H$ is
now some rotationally invariant function of the invariants (\ref{inv1}). Any $H$ which is a function of (\ref{inv1}), and rotationally invariant, can in
principle be used, giving rise to a fairly rich spectrum of possible interacting Carroll particles.  This does not contradict
the representation theory as that only applies to the center of mass degree of freedom. These examples were studied in \cite{Bergshoeff:2014jla}.

It is also possible to take the other point of view where we for example impose first class constraints $E_a=0$ for all $a$ and add an explicit Hamiltonian $H$.
But now something interesting happens. As long as the Hamiltonian does not involve the time coordinates, its Poisson bracket
with the generator of Carroll boosts will be proportional to a linear combination of the $E_a$. This is not zero, but we can compensate for this by a suitable
transformation rule for the Lagrange multipliers. Therefore, we need not worry about boost invariance of $H$, and in fact any $H$
which preserves translations and rotations is admissible. This is no different from standard many-particle dynamics. It looks like
any many-particle Hamiltonian can be dressed with a Carroll symmetric center of mass degree of freedom to provide a Carroll
invariant system of particles each with zero energy. This case was not considered in \cite{Bergshoeff:2014jla}, and it might be instructive to get these
systems directly as a $c\rightarrow 0$ limit of a relativistic many-particle system.

\section{Carroll Field Theory \label{sec:CFT} }

In this section, we switch to field theory, and prepare for the application to cosmology we have in mind. The starting point is to construct Lagrangians that are Carroll invariant and can be obtained from a $c\to 0$ limit of a relativistic theory or through an expansion in small $c$. We treat here the real scalar field and the Maxwell field as examples. 

\subsection{Scalar field}\label{ZEL}
Consider a relativistic real scalar field $\phi$. Under a Lorentz boost, it transforms as
\begin{equation}
    \delta\phi=ct \vec\beta\cdot\vec\partial\phi+\frac{1}{c}\vec\beta\cdot\vec x \,\partial_t\phi\ .
\end{equation}
The relativistic scalar field Lagrangian density
\begin{equation}\label{eq:scalarcarroll11}
    \mathcal{L}
    =
    \frac{1}{2c^{2}}
    \left(\partial_{t}\phi\right)^{2}
    -
    \frac{1}{2}
    \left(
        \partial_{i}\phi
    \right)^{2}
    -
    V(\phi)
    \,.
\end{equation}
transforms into a total derivative, as is well known. The conjugate momentum is 
\begin{equation}
    \pi_\phi=\frac{1}{c^2}\partial_t\phi\ .
\end{equation}
The relativistic energy-momentum tensor is
\begin{equation}\label{scalarstresstensor}
T^\mu{}_\nu=\delta^\mu{_\nu}\left( -\frac{1}{2}\partial_\rho\phi\partial^\rho\phi-V(\phi)\right)+\partial^\mu\phi \partial_\nu\phi\ ,
\end{equation}
and is symmetric when raising the indices.

For a quadratic potential of the form $V(\phi)=\frac{1}{2}\frac{m^2c^2}{\hbar^2}\phi^2$ 
(in SI units), we get the usual relativistic dispersion relation from a plane wave ansatz $\phi\propto e^{i\vec{k}\cdot\vec{x}+i\omega t}$,
\begin{equation}\label{reldisprel}
    E^2\equiv \hbar^2\omega^2=c^2\vec{p}^{\,2}+m^2c^4\ ,
\end{equation}
with $\vec p=\hbar \vec k$.

We now make an expansion around $c=0$ so we write\footnote{Odd powers of $c$ do not play a role in all examples we consider.} 
\begin{equation}\label{exp_phi}
    \phi=c^{\Delta}\Big(\phi_0+c^2\phi_1+c^4\phi_2+\cdots\Big)=c^{\Delta}\sum_{n=0}^{\infty}\phi_nc^{2n}\ ,
    \end{equation}
for some $\Delta$.
Defining as before $\vec\beta=c\vec b$, with $\vec b$ the Carroll boost parameter, one finds the Carroll transformations\footnote{In principle, one could generate an algebra expansion by writing $\vec\beta=c[\vec b_0+c^2\vec b_1+...]$ where at each order, there is a new symmetry generated by $\vec b_n$, but we simply restrict here to the leading order Carroll algebra, in which only $\vec b_0$ is non-zero. For example, at the next-to-leading order the symmetry transformation corresponding to $b_1$ is $\delta\phi_1=\vec b_1\cdot \vec x\partial_t\phi_0$ and $\delta\phi_0=0$.}, for $n>0$,
    \begin{equation}
\delta\phi_0=\vec b\cdot \vec x \,\partial_t \phi_0\ ,\qquad        \delta\phi_n=\vec b\cdot \vec x \,\partial_t \phi_n +t\, \vec b \cdot \vec \partial \phi_{n-1}\ .
    \end{equation}
The field $\phi_0$ is a scalar field with respect to Carroll transformations, as the Carroll generator for boosts is $C_i=x^i\partial_t$. The higher order modes in the expansion are not Carroll scalars, and transform into each other under boosts. 
    
Using this expansion, it is rather straightforward to construct Carroll invariant Lagrangians. For free fields, we find, for the two lowest orders in the small-$c$ expansion,
\begin{equation}
{\cal L}_0=\frac{1}{2}\dot\phi_0^2\ ,\qquad {\cal L}_1=\dot\phi_0\dot\phi_1-\frac{1}{2}\partial_i\phi_0\partial_i\phi_0\ ,
\end{equation}
where we defined ${\cal L}_0$ and ${\cal L}_1$ via ${\cal L}=c^{2\Delta-2}\left({\cal L}_0+c^2{\cal L}_1+O(c^4)\right)$.
One can explicitly check that both ${\cal L}_0$ and ${\cal L}_1$ transform into total derivatives so their corresponding actions are invariant with the appropriate boundary conditions. Note that $\mathcal{L}_{1}$ contains the field $\phi_{1}$ which does not transform as a scalar field under Carroll transformations. As we discuss in more detail below ${\cal L}_1$ is not Carroll boost invariant but can be made to be by adding a Lagrange multiplier that sets $\dot\phi_0$ equal to zero on shell.

It is easy to add interactions, starting from a Lorentz invariant potential by simply substituting the expansion \eqref{exp_phi} in the potential. Different choices can be made however, depending on how the coupling constants in the potential scale with $c$. We consider here the simplest example of a quadratic potential. Assuming the parameters depend homogeneously on $c$ there are still two choices, namely keeping $E_0\equiv mc^2$ (or $\omega_0=E/\hbar$) fixed, or keeping the Compton wavelength $\lambda^{-1}=\mu\equiv mc/\hbar$ fixed, similar to the analysis of the relativistic particle in Section \ref{tachyonCarroll}.  

For fixed $\omega_0$, we get 
\begin{eqnarray}\label{L0L1}
{\cal L}_0&=&\frac{1}{2}\dot\phi_0^2-\frac{1}{2}\omega_0^2\phi_0^2\ ,\nonumber\\
{\cal L}_1&=&\dot\phi_0\dot\phi_1-\frac{1}{2}\partial_i\phi_0\partial_i\phi_0 - \omega_0^2\phi_0\phi_1\ .
\end{eqnarray}
The equation of motion associated to ${\cal L}_0$ is
\begin{equation}\label{EOMphi0_omega}
    \ddot\phi_0+\omega_0^2\phi_0=0\ .
\end{equation}
For ${\cal L}_1$ the equation of motion for $\phi_1$ is the same as in \eqref{EOMphi0_omega}{}\footnote{This is a special case of a general result that states that the equations of motion of a Lagrangian at order $N$, say, contain all the equations of motion of the Lagrangians at orders $n<N$ via the dependence of the $N$th order Lagrangian on the subleading fields, see \cite{Hansen:2020pqs} (section 2.5).}, but we get a second equation by varying $\phi_0$,
\begin{equation}
    \ddot\phi_1+\omega_0^2\phi_1=\partial^2\phi_0\ .
\end{equation}
The solutions to the response of $\phi_{1}$ to $\phi_{0}$ can be found in terms of plane waves,
\begin{equation}\label{phi0phi1}
    \phi_0=e^{i\vec k\cdot \vec x+i\omega_0 t}+c.c.\ ,\qquad \phi_1=\frac{i}{2\omega_0}{\vec k}^{\,2}\,t\,e^{i\vec k\cdot \vec x+i\omega_0 t}+c.c.
    \,.
\end{equation}
Both $\phi_0$ and $\phi_1$ are fields arising in the expansion around $c=0$, and we can use them to reconstruct the relativistic scalar using \eqref{exp_phi}: $\phi=\phi_0+c^2\phi_1+\cdots$, where we put $\Delta=0$ for simplicity. Then the full relativistic plane wave solutions are known of course, and involve the frequency in \eqref{reldisprel}. We can expand the relativistic dispersion relation around $c=0$ using
\begin{equation}
    \omega=\omega_0+c^2\frac{{\vec k}^{\,2}}{2\omega_0}+\cdots\ ,\qquad e^{i\omega t}=e^{i\omega_0 t}\Big(1+i\,c^2 \frac{{\vec k}^{\,2}}{2\omega_0}\,t+\cdots \Big)\ ,
\end{equation}
where we remind that $\omega_0=mc^2/\hbar$ which we kept fixed in the Carroll limit. One now sees that this expansion is 
reproducing the expansion in $\phi$ using \eqref{phi0phi1}, as it should. Furthermore, if we perform an infinitesimal Carroll transformation on the right hand side of the dispersion relation, we find $\delta\omega=-c^2\vec k\cdot \vec b=-c\vec k\cdot\vec\beta$, which is the first term in the expression for an infinitesimal Lorentz transformation on the energy or frequency. If we keep expanding further in the $\phi_n$, we will reconstruct the full relativistic solution.

Now we consider the second possibility, in which we keep $\mu=mc/\hbar$ fixed in the Carroll limit. Then we find
\begin{eqnarray}
{\cal L}_0&=&\frac{1}{2}\dot\phi_0^2\ ,\nonumber\\
{\cal L}_1&=&\dot\phi_0\dot\phi_1-\frac{1}{2}\partial_i\phi_0\partial_i\phi_0 -\frac{1}{2} \mu^2\phi_0^2\ .
\end{eqnarray}
The equations of motion for ${\cal L}_0$ is
\begin{equation}\label{EOMphi0}
    \ddot\phi_0=0 \qquad \Rightarrow \qquad \phi_0=f(\vec x)+g(\vec x)t\ .
\end{equation}
For ${\cal L}_1$ the equation of motion for $\phi_1$ is the same as in \eqref{EOMphi0}, but we get a second equation
\begin{equation}
    \ddot\phi_1=\partial^2\phi_0-\mu^2\phi_0\ .
\end{equation}
If $\ddot\phi_1=0$, and upon taking plane spatial waves for $\phi_0$ (i.e. $f(\vec x)=e^{i\vec k\cdot \vec x}+c.c.$), one arrives at the tachyonic modes for a single particle again, with $\vec k^{\,2}+\mu^2=0$. However, the interaction between $\phi_0$ and $\phi_1$ provides non-trivial solutions with $\vec k^{\,2}+\mu^2>0$. 

The relativistic dispersion relation in this case is written as $\omega=\pm c{\sqrt{\vec k^{\,2}+\mu^2}}$ and vanishes in the strict Carroll limit. This corresponds to Carroll representations with vanishing energy, $E=0$. The first correction is linear in $c$ and gives already the exact result for $\omega$. 

To conclude this part of the discussion, we find again, just as for the relativistic particle, two representations in the Carroll limit, those with non-zero energy $E_0=\hbar\omega_0=mc^2$ and those with zero energy where we kept $mc/\hbar$ fixed.

We now consider some further properties of ${\cal L}_0$ and ${\cal L}_1$.

\subsubsection{Energy-momentum tensor}\label{sectL0}

We start with the  Carroll invariant Lagrangian
\begin{equation}
    {\cal L}_0=\frac{1}{2}\dot\phi_0^2-V(\phi_0)\ .
\end{equation}
The corresponding energy-momentum tensor is\footnote{We compute the energy-momentum tensor from the Noether procedure, 
\begin{equation}\label{TNoether}
    T^\mu{}_\nu=\delta^\mu{}_\nu{\cal L}-\frac{\partial {\cal L}}{\partial(\partial_\mu\phi)}\partial_\nu\phi\ .
\end{equation}},
\begin{equation}\begin{aligned}\label{Eq:EMT-L0}
    T^{t}{_t}
    &=-\Big(
    \frac{1}{2}\dot\phi_0^2+V\Big)
    \,,
    \quad
    T^{i}{_t}=0
    \,,
    \\
    T^{t}{_i}
    &
    =
    -\dot\phi_0 \partial_{i}\phi_0
    \,,
    \quad
    T^{i}{_j}
    =
    \left(
        \frac{1}{2}\dot\phi_0^{2}-V
    \right)\delta^{i}{_j}
    \,.
\end{aligned}\end{equation}
This energy-momentum tensor obeys the Carroll transformation laws \eqref{eq:trafolaws} and the constraint that $T^i{}_t=0$. Notice that spatial derivatives $\partial_i\phi$ are absent in ${\cal L}_0$ but they do appear in the energy-momentum tensor. Their appearance is important for showing that the energy-momentum tensor is conserved on shell. Notice furthermore that this energy-momentum tensor arises as the leading order term in the expansion
\begin{equation}
\label{Texpansion} 
    T^\mu{}_\nu=T_{(0)}^\mu{}_\nu+c^2 T_{(1)}^\mu{}_\nu+\cdots\ , 
\end{equation}
where the left-hand side stands for the relativistic energy-momentum tensor which transforms as a tensor under Lorentz transformations. On the right-hand side, the leading term $T_{(0)}^\mu{}_\nu$ transforms as a Carroll tensor and in particular it should obey $T^i{}_t=0$. The first order correction to it, $T_{(1)}$ does not transform as a tensor under the Carroll group (it will transform into $T_{(0)}$ as well), just like $\phi_1$ does not transform as a scalar under Carroll boosts. Therefore $T_{(1)}$ is not expected to satisfy the Carroll identity $T^i{}_t=0$, as we will explicitly see below.

The theory described by the next-to-leading order action has the Lagrangian 
\begin{equation}\label{L1}
    {\cal L}_1=\dot\phi_0\dot\phi_1-\frac{1}{2}\partial_i\phi_0\partial_i\phi_0\ ,
\end{equation}
where we ignored possible potential terms.
 Using \eqref{TNoether}, adapted to a theory containing two scalar fields, the energy-momentum tensor components now read 
\begin{eqnarray}
&&T^t{}_t=-\dot\phi_0\dot\phi_1-\frac{1}{2}(\partial_i\phi_0)^2\ ,\qquad T^i{}_t=\dot\phi_0\partial_i\phi_0\ ,\nonumber\\
&&T^t{}_i=-\dot\phi_1\partial_i\phi_0-\dot\phi_0\partial_i\phi_1\ ,\qquad T^i{}_j=\delta^i{_j}{\cal L}_1+\partial^i\phi_0\partial_j\phi_0\ .
\end{eqnarray}
One can check that it is conserved, but it does not  satisfy the Carroll constraint $T^i{}_t=0$, as explained above. One can also check that it follows from the relativistic symmetric energy-momentum tensor by expanding in powers of $c$, i.e. the $T_{(1)}^\mu{}_\nu$ term in \eqref{Texpansion}.

The expansion of the relativistic scalar field Lagrangian around $c=0$ leads to (after appropriately rescaling the Lagrangian with $c^{2-2\Delta}$) the Lagrangians $\mathcal{L}_0$ and $\mathcal{L}_1$. 
The Lagrangian $\mathcal{L}_0$ is Carroll boost invariant while the Lagrangian $\mathcal{L}_1$ is not\footnote{The next-to-leading order Lagrangian is invariant under a symmetry group whose Lie algebra can be obtained by expanding the Poincar\'e algebra around $c^2=0$ and quotienting this algebra by keeping only the level zero and level one generators (see \cite{dBHOSV3} for more details).}. We can however modify $\mathcal{L}_1$ by adding a Lagrange multiplier to make it Carroll boost invariant. The Lagrange multiplier sets to zero $\dot\phi_0$ (so that now $\mathcal{L}_1$ is the leading order Lagrangian in the expansion of the free scalar theory). This leads to 
\begin{equation}
    {\cal L}=\dot\phi_0\dot\phi_1-\frac{1}{2}\partial_i\phi_0\partial_i\phi_0+\tilde\chi\dot\phi_0\ ,
\end{equation}
with Lagrange multiplier $\tilde\chi$.
We can absorb $\dot\phi_1$ into $\tilde\chi$ leading to 
\begin{equation}
    {\cal L}=-\frac{1}{2}\partial_i\phi_0\partial_i\phi_0+\chi\dot\phi_0\ ,
\end{equation}
where the new Lagrange multiplier is $\chi$.

We can obtain the latter Lagrangian also directly by taking a Carroll limit. Let us consider the following relativistic Lagrangian density consisting of two scalar fields $\chi$ and $\phi$:
\begin{equation}\label{eq:scalarcarroll1}
    \mathcal{L}
    =
    -
    \frac{c^{2}}{2}\chi^{2}
    +
    \chi \partial_{t}\phi
    -
    \frac{1}{2}
    \left(
        \partial_{i}\phi
    \right)^{2}
    -
    V(\phi)
    \,.
\end{equation}
From the equation of motion for $\chi$,
\begin{equation}\label{expr_chi}
    \chi
    =
    \frac{1}{c^{2}}\partial_{t}\phi
    \,,
\end{equation}
it is easy to see that the Lagrangian density in \eqref{eq:scalarcarroll1} is engineered such that integrating out $\chi$ yields a canonical relativistic scalar field Lagrangian as given in \eqref{eq:scalarcarroll11}. And of course, $\chi$ is identical to the conjugated momentum when seen in first order formalism. It is however not a scalar field under Lorentz transformations.


In the Carroll limit $c\to 0$ keeping both $\chi$ and $\phi$ fixed, we find that the Lagrangian density in \eqref{eq:scalarcarroll1} becomes 
\begin{equation}\label{eq:Carrollscalarchi}
    \mathcal{L}
    =
    \chi \dot\phi
    -
    \frac{1}{2}\left(
        \partial_{i}\phi
    \right)^{2}
    -
    {V}(\phi)
    \,.
\end{equation}
The action is invariant under Carroll boosts if we assign the transformation laws 

\begin{equation}
    \delta\phi=\vec b\cdot \vec x \,\dot\phi\ ,\qquad \delta\chi=\vec b\cdot \vec x \,\dot\chi + \vec b\cdot \vec \partial \phi\ .
\end{equation}

The field equation of $\chi$ sets $\dot{\phi}=0$, in fact $\chi$ is a Lagrange multiplier. This is consistent with the expression \eqref{expr_chi} and the fact that both $\chi$ and $\phi$ are kept fixed in the $c\to 0$ limit. The situation is then as described in the introduction, and in the small $c$-expansion, this means that the scalar field should be slowly varying in time.

Notice that the Lagrangian \eqref{eq:Carrollscalarchi} is actually very similar to \eqref{L1}. It seems that all one needs to do is to perform the field redefinition $\chi=\dot\phi_1$. However this changes the Lagrangian in a non-trivial way as varying $\phi_1$ does not lead to the same equation of motion as varying $\chi$. Using the on-shell constraint $\dot{\phi}=0$, the resulting energy-momentum tensor is now given by

\begin{equation}\begin{aligned}\label{eq:L3}
    T^{t}{_t}
    =&~
    -\left({V}+\frac{1}{2}(\partial_{i}\phi)^{2}\right)
    \,,
    \quad
    T^{i}{_t}=~
    0\,,
    \\
    T^{t}{_j}
    =&~
    -\chi\partial_{j}\phi
    \,,\quad
    T^{i}{_j}
    =~
    -\left(
    {V}
    +
    \frac{1}{2}(\partial_{i}\phi)^{2}
    \right)\delta^{i}{_j}
    +
    \partial_{i}\phi
    \partial_{j}\phi
    \,.
\end{aligned}\end{equation}
It is conserved on-shell and indeed, as required by Carroll symmetry, $T^{i}{_t}=0$. We see now that if we subject $\mathcal{L}_{1}$ in \eqref{L1} to $\dot{\phi}_{0}=0$, that the two theories are equivalent when $\chi$ plays the role of $\dot{\phi}_{1}$ and the energy-momentum tensor in \eqref{L1} becomes the same as \eqref{eq:L3}. 

\subsection{Maxwell theory}
 We can also consider Carrollian versions of Maxwell's theory.
In analogy to the scalar field theory case treated in the previous section, we present here again
two perspectives. One in which we consider an expansion for slow speed of light and the other based on taking the $c \rightarrow 0$ limit, as has been studied in e.g. \cite{Duval:2014uoa,Bagchi:2019xfx,Basu:2018dub,Bagchi:2019clu,Banerjee:2020qjj}. Furthermore, in parallel with similar investigations
 of the  non-relativistic expansion/limit ($c \rightarrow \infty$) of Maxwell theory (see e.g. \cite{Festuccia:2016caf,Festuccia:2016awg}), 
one finds two distinct sectors, the electric and the magnetic.


\subsubsection{Expansions}

The Maxwell field $A_\mu$ transforms under Lorentz boosts as
\begin{equation}
    \delta A_\mu =ct  \vec\beta\cdot\vec\partial A_\mu +\frac{1}{c}\vec\beta\cdot\vec x \,\partial_t A_\mu + \bar \delta A_\mu
    \,,
\end{equation}
where $\bar \delta A_0 =   \vec\beta \cdot \vec{A} $ and $\bar \delta \vec{A} =  \vec\beta A_0 $. Here we have used the transformation $\delta A_\mu = \xi^{\nu}\partial_{\nu}A_{\mu}+\partial_{\mu}\xi^{\nu}A_{\nu}$ under a general coordinate transformation
and used its restriction to a Lorentz transformation via $\xi^0 = \vec{\beta} \cdot \vec{x}/c$, $\xi^i = x^0/c \beta^i = t \beta^i $.
From now on we will use 
  $A_t = c A_0$, which allows%
  \footnote{This is also natural when considering the 1-form $A = A_\mu dx^\mu = A_0 dx^0 + A_i dx^i = 
  A_t dt + A_i dx^i$, while it also ensures we can write $F_{i0} = \partial_i A_0 - \partial_0 A_i =
 \frac{1}{c} (\partial_i A_t - \partial_t A_i)= \frac{E_i}{c}$.}
    for an expansion in even powers of $c$ such that
 \begin{equation}\label{exp_A}
    A_t =c^{\Delta}\Big(
    A_t^{(0)} + c^2 A_t^{(1)} + \cdots\Big)=c^{\Delta}\sum_{n=0}^{\infty} A_t^{(n)}c^{2n}\ ,
    \end{equation}
    for some $\Delta$, and likewise for $A_i$ with the same $\Delta$. 
    
Again we define $\vec\beta=c\vec b$, with $\vec b$ the Carroll boost parameter. The fields in the expansion above then transform as 
    \begin{equation}
    \label{Attrafo} 
\delta A_t^{(n)} =\vec b\cdot \vec x \,\partial_t A_t^{(n)}
+t\, \vec b \cdot \vec \partial A_t^{(n-1)} + \vec b \cdot 
\vec A^{(n-1)} \ ,
\end{equation}
\begin{equation}
\label{Aitrafo} 
    \delta A_i^{(n)} =\vec b\cdot \vec x \,\partial_t A_i^{(n)}
    + b_i A_t^{(n)} + t\, \vec b \cdot \vec \partial A_i^{(n-1)} \ ,
\end{equation}
where the transformations for $n=0$ are included using $A_\mu^{(-1)}\equiv 0$.  The field 
$(A_t^{(0)},A_i^{(0)})$ is a vector field
with respect to Carroll transformations. 
 The higher order modes in the expansion  transform into each other under boosts. 

Next we study the expansion of the Maxwell Lagrangian 
\begin{equation}
        \mathcal{L} = 
    \frac{1}{2c^{2}}(E^{i})^{2}
    -
    \frac{1}{4}(F_{ij})^{2} \ ,
\end{equation}
where the field strengths are
\begin{equation}
    E_i =   \partial_i A_t - \partial_t A_i \ ,
\end{equation}
\begin{equation}
     F_{ij} = \partial_i A_j -  \partial_j A_i \ . 
\end{equation}

Inserting the expansion of the vector field we thus find that the Lagrangian expands according
to ${\cal{L}} = c^{2 \Delta -2} (   {\cal{L}}_0 + c^2  {\cal{L}}_1 + \cdots)$, where the leading order Lagrangian
is 
\begin{equation}
    {\cal{L}}_0 = \frac{1}{2} (E_i^{(0)} )^2 \ ,
\end{equation}
with 
\begin{equation}
    E_i^{(0)} =   \partial_i A_t^{(0)} - \partial_t A_i^{(0)} \ .
\end{equation}
The next to leading order Lagrangian is 
\begin{equation}\label{eq:lagrangian1}
    {\cal{L}}_1 =  E_i^{(0)} E_i^{(1)} - \frac{1}{4} (F_{ij}^{(0)})^2 \ ,
\end{equation}
where
\begin{equation}
    F_{ij}^{(0)} = \partial_i A_j^{(0)} - \partial_j A_i^{(0)} \ ,
\end{equation}
\begin{equation}
    E_i^{(1)} =\partial_i A_t^{(1)}  -  \partial_t A_i^{(1)} \ . 
\end{equation}
One can explicitly check that  ${\cal{L}}_0$ and ${\cal{L}}_1$ are invariant under the transformations
in \eqref{Attrafo}, \eqref{Aitrafo}, up to total derivatives. 

Finally, the Bianchi identity $\epsilon^{\alpha\lambda\mu\nu}\partial_{\lambda}F_{\mu\nu}=0$ implies
\begin{equation}\label{eq:bianchi1}
	\partial_{i}B_{i}^{(n)}
	=
	0
	\,,
	\quad
	\partial_{t}B_{i}^{(n)}
	+
	(\nabla\times E^{(n)})_{i}
	=
	0
	\,,
\end{equation}
where $ B_i^{(n)} = \frac{1}{2} \epsilon_{ijk}  F^{(n)}_{jk}$. 
\subsubsection{Electric Carroll sector}
Let us first focus on  ${\cal{L}}_0$. This is the action of the electric sector of Carrollian electrodynamics
previously identified in \cite{Duval:2014uoa,Bagchi:2014ysa}. 
It also follows directly from the strict $c \rightarrow 0$ limit of the Maxwell action.  
To avoid clutter, we drop the 0-superscript on all the fields ($A_t^{(0)}$, $A_i^{(0)}$, etc.) in this specific subsection.
The corresponding equations of motion are 
\begin{equation}\label{eq:electricmaxwell1}
 \partial_i E_i = 0 \, , \quad \partial_t E_i  = 0\,, 
\end{equation}
which are respectively Gauss' law and Amp\`ere's law with the Amp\`ere term switched off, as is known
for the electric limit of Carroll electromagnetism \cite{Duval:2014uoa,Bagchi:2014ysa}. In addition we have the Bianchi identities \eqref{eq:bianchi1} giving
\begin{equation}\label{eq:electricmaxwell2}
 \partial_i B_i = 0 \, , \quad \partial_t B_i  + (\vec\nabla\times\vec E)_{i}   = 0 \,.
\end{equation}
Furthermore, using the transformation rules of $(A_t^{(0)}, A_i^{(0)})$ in \eqref{Attrafo}, \eqref{Aitrafo},
we find the Carroll transformations
\begin{equation}
\delta E_i  = \vec b\cdot \vec x \,\partial_t E_i  \, , \quad 
\delta B_i  = \vec b\cdot \vec x \,\partial_t B_i - (\vec b \times \vec E)_i\,,
\end{equation}
which indeed leave the equations of motion of the electric Carrollian sector invariant. 
The corresponding energy-momentum tensor is given by
\begin{equation}\label{eq:emtE}\begin{aligned}
    T^{t}{_t}
    =&~
    -\frac{1}{2}(E^{i})^{2}
    \,,
    \quad
    T^{i}{_t}
    =~
    0
    \,,
    \\
    T^{t}{_j}
    =&~
	(\vec E\times\vec B)_{j}
    \,,\quad
    T^{i}{_j}
    =~
    -E^{i}E_{j}+\frac{1}{2}(E^{i})^{2}\delta^{i}{_j}
    \,,
\end{aligned}\end{equation}
where we used the following improved formula for the energy-momentum tensor such that $T^{i}{_t}=0$ and $T^{i}{_j}=T^{j}{_i}$
\begin{equation}
    T^{\mu}{_\nu}
    =
    -
    \frac{\delta\mathcal{L}}{\delta\partial_{\mu}A_{\alpha}}\partial_{\nu}A_{\alpha}
    +
    \delta^{\mu}{_\nu}\mathcal{L}
    -
    \left(
        \delta^{\mu}{_k}\partial_{t}
        -
        \delta^{\mu}{_t}\partial_{k}
    \right)(E^{k}A_{\nu})
    \,.
\end{equation}
The resulting energy-momentum tensor is Carroll covariant and remains traceless in 3+1 dimensions. This result coincides with what one would get from taking the non-relativistic limit of the Lorentzian energy-momentum tensor.

Let us consider general solutions to the electric Carroll sector. From \eqref{eq:electricmaxwell1} we find
\begin{equation}
    \vec E
    =
    \vec{\nabla}\times\vec{V}(\vec{x})
    \,,
\end{equation}
where $\vec{V}(x,y,z)$ is some arbitrary vector field. We see that the electric field is static. From the second equation in \eqref{eq:electricmaxwell2} we obtain the solution
\begin{equation}
    B_{i}
    =
    \psi_{i}(\vec{x})
    -
    t(\vec{\nabla}\times \vec{E})_{i}
    \,.
\end{equation}
From the first condition in \eqref{eq:electricmaxwell2} we obtain the following solution for $\psi_{i}$: $\partial_{i}\psi_{i}(\vec{x})=0$. In contrast to the electric field, the magnetic field is found to allow for time dependence, be it only linearly.

Let us consider again the electric Carroll equations of motion given by equations \eqref{eq:electricmaxwell1} and \eqref{eq:electricmaxwell2}, but this time in momentum space. These imply that $\partial_t^2 \vec B=0$ so that $\vec B=\vec B_0+t\vec B_1$. The fields $\vec E$, $\vec B_0$ and $\vec B_1$ are time-independent and they obey
\begin{equation}
    \vec\nabla\cdot \vec E=0\,,\qquad\vec\nabla\cdot \vec B_0=0\,,\qquad\vec\nabla\times\vec E+\vec B_1=0\,.
\end{equation}
If we insert a plane wave profile $\vec E=\vec{\mathcal{E}} e^{i\vec k\cdot \vec x}$, $\vec B_0=\vec{\mathcal{B}}_0 e^{i\vec k\cdot \vec x}$ and $\vec B_1=\vec{\mathcal{B}}_1 e^{i\vec k\cdot \vec x}$ this leads to the algebraic equations
\begin{equation}
    \vec k\cdot \vec{\mathcal{E}} =0\,,\qquad\vec k\cdot \vec{\mathcal{B}}_0=0\,,\qquad i\vec k\times\vec {\mathcal{E}}+\vec{\mathcal{B}}_1=0\,.
\end{equation}
It follows that $\vec k$, $\vec{\mathcal{E}}$ and $\vec{\mathcal{B}}_1$ form an orthogonal system. Hence given $\vec k$ and, say $\vec{\mathcal{E}}$, we know $\vec{\mathcal{B}}_1$. The only freedom left is a rotation of the orthogonal pair $\vec{\mathcal{E}}$ and $\vec{\mathcal{B}}_1$ around the momentum $\vec k$. This freedom is a manifestation of the helicity discussed earlier in Section \ref{subsec:rep} on representations of the Carroll algebra. A similar analysis goes through for the magnetic limit.

\subsubsection{Magnetic Carroll sector} 


Next we turn to  ${\cal{L}}_1$. As we will shortly see this is closely related to a (novel) action for
the magnetic sector of Carrollian electrodynamics.
First of all, the equations of motion obtained by varying $(A_t^{(1)}, A_i^{(1)})$ will give rise
to the equations of motion of ${\cal{L}}_0$, which is a general feature of expanded Lagrangians. 
Furthermore, varying the leading order fields gives the equations
\begin{equation}
 \partial_i E_i^{(1)} = 0 \, , \quad  - \partial_t E_i^{(1)}  +(\nabla\times B^{(0)})_{i}=0
\end{equation}
which exhibit respectively Gauss law and Amp\`ere's law, but with the electric field
 $E_i^{(1)} $  (instead of  $E_i^{(0)}$). In addition coming from \eqref{eq:bianchi1} we have the Bianchi identity 
 \begin{equation}
 \label{eq:Bianch} 
 	\partial_i  B_i^{(0)}=0
	\,,\quad
 	\partial_t  B_i^{(0)}
	+
	(\nabla\times E^{(0)})_{i}
	=0
	\,.
\end{equation}
Let us consider the case when we force the leading term in the expanded action to be zero,  i.e.  $E_i^{(0)}=0$. We can achieve this by adding a Lagrange multiplier term $\tilde\chi_i E_i^{(0)}$ to ${\cal L}_1$ with Lagrange multiplier $\tilde\chi_i$.
In that case we thus also have from \eqref{eq:Bianch}  that $ \partial_t B_i^{(0)} = 0$ which is Faraday's law with only the magnetic  induction term. 
 Thus we recover the equations of motion of the magnetic Carroll limit \cite{Duval:2014uoa,Bagchi:2014ysa}.  As a result the action  ${\cal{L}}_1$ (which is now the leading term) becomes a Carroll invariant action for the magnetic sector, something which was previously unknown. 
As a further check we compute the Carroll transformations of the fields
\begin{equation}
\label{E1trafo} 
\delta E_i^{(1)} = \vec b\cdot \vec x \,\partial_t E_i^{(1)}  + (\vec b \times \vec B^{(0)})_i \, , \quad 
\delta B_i^{(0)} = \vec b\cdot \vec x \,\partial_t B_i^{(0)}  \,,
\end{equation}
which indeed leave the equations of motion of the magnetic Carrollian sector invariant. 

We can also find the action for the magnetic Carroll sector using a limiting procedure analogous to the one
considered for the scalar field. For this we first  introduce an altered Maxwell action containing a new field $\chi^{i}$ which when integrated out yields the original Maxwell action, i.e.
\begin{equation}
    \mathcal{L}
    =
    -\frac{c^{2}}{2}\chi_{i}\chi_{i}
    +
    \chi_{i}E_{i}
    -
    \frac{1}{4}
    (F_{ij})^{2}
    \,.
\end{equation}
Here $\chi^{i}$ has to transform in a particular manner such that the Lagrangian remains invariant under Lorentz boosts.
Now taking $c\to0$ 
we obtain Carroll boost invariant Lagrangian
\begin{equation}\label{eq:maxwell4}
    \mathcal{L}
    =
    \chi_{i}E_{i}
    -
    \frac{1}{4}
    (F_{ij})^{2}
    \,.
\end{equation}
Here the magnetic part of the Maxwell tensor survives the Carroll limit. 
Furthermore, one has to require the following transformation of $\chi_{i}$ under Carroll boost: \begin{equation}
\label{chitrafo} 
    \delta \chi_{j}
    =
    \vec{b}\cdot \vec{x}\partial_{t}\chi_{j}
    +
    \left(
    	\vec{b}\times \vec{B}
    \right)_{j}
    \,,
\end{equation}
in order to keep the Lagrangian invariant under the Carroll boost. 
Note that the Lagrangian is precisely of the form of ${\cal{L}}_1$ in \eqref{eq:lagrangian1} with the difference
that instead of $E_i^{(1)}$ we have the Lagrange mutiplier $\chi_i$. (cf. $\dot \phi_1$ vs. $\chi$ for the
scalar case). 

The equations of motion are easily computed. First of all the Lagrange multiplier $\chi_i$ enforces
$E_i=\partial_i A_t-\partial_t A_i=0$ and hence $\partial_t F_{ij} =0$. The remaining equations of motion are then 
\begin{equation}
\partial_i \chi_i = 0 \, , \quad 
 -\partial_{t}\chi_{i}
 +
 (\vec{\nabla}\times \vec{B})_{i}
 =0
 \,.
\end{equation}
Together with the Bianchi identity for the $B$-field,
 \begin{equation}
 	\partial_i  B_i=0
	\,,\quad
 	\partial_t  B_i
	=0
	\,,
\end{equation} 
we thus see that these are the correction equations of motion
for magnetic Carroll upon identifying $\chi_i$ with the (true) electric field ($E_i^{(1)}$ in the analysis above). Indeed the transformation of $\chi_i$ in  \eqref{chitrafo} correctly corresponds to the transformation of $E_i^{(1)} $ in 
 \eqref{E1trafo}.

We also give here the resulting energy-momentum tensor of the novel action \eqref{eq:maxwell4} for magnetic Carroll.
This takes the form 
\begin{equation}\label{eq:emtB}\begin{aligned}
    T^{t}{_t}
    =&~
	-\frac{1}{2}(B_{i})^{2}
    \,,
    \quad
    T^{i}{_t}
    =~
    0
    \,,
    \\
    T^{t}{_j}
    =&~
    (\vec{\chi}\times \vec{B})_{j}
    \,,\quad
    T^{i}{_j}
    =~
   -B_{i}B_{j}+\frac{1}{2}(B_{k})^{2}\delta^{i}{_j}
    \,,
\end{aligned}\end{equation}
upon using $E_{i}=0$ as well as the improved energy-momentum tensor formula
\begin{equation}
    T^{\mu}{_\nu}
    =
    -
    \frac{\delta\mathcal{L}}{\delta\partial_{\mu}A_{\alpha}}\partial_{\nu}A_{\alpha}
    +
    \delta^{\mu}{_\nu}\mathcal{L}
    -
    \delta^{\mu}{_t}
    \partial_{i}
    \left[
        \chi_{i}A_{\nu}
    \right]
    +
    \delta^{\mu}{_i}
    \left[
        \partial_{t}(\chi_{i}A_{\nu})
        +
        \partial_{j}(F_{ij}A_{\nu})
    \right]
    \,.
\end{equation}
This energy-momentum tensor remains traceless in 3+1 dimensions and correctly satisfies $ T^{i}{_t} =0 $. 
It was noted in \cite{Duval:2014uoa} that the electric and magnetic Carroll sectors are related via electromagnetic duality
which acts as $ \vec E \rightarrow \vec B$, $\vec B \rightarrow -\vec{E}$,  
as opposed to relativistic Maxwell which is invariant. This is obvious from the equations of motion. Comparing
the energy momentum tensors in \eqref{eq:emtE} and \eqref{eq:emtB} we see that these also respect this symmetry.
Note in particular that the momentum current is the Poynting vector $\vec E \times \vec B$ which is
thus invariant. Finally, using the electromagnetic duality we can also recycle the solutions to the equations of motion obtained in the electric sector for the magnetic sector.

Another way to view both sectors is by starting from a relativistic Maxwell energy-momentum tensor and considering the following dimensionless combinations
\begin{equation}
    \frac{|B|}{(|E|/c)}
    =c\left/
    \left(
    \frac{|E/c|^{2}}{|\mathcal{P}|}
    \right)
    \right.
    \ll1\quad \text{(electric)}
\,,
    \quad
    \frac{(|E|/c)}{|B|}
    =
    c
    \left/
    \left(
    \frac{|B|^{2}}{|\mathcal{P}|}
    \right)
    \right.
    \ll1\quad \text{(magnetic)}
    \,,
\end{equation}
while keeping the relativistic momentum density flux $|\mathcal{P}|=|E||B| c^{-2}$ fixed and requiring that  $|E/c|$ is fixed for the electric case or $|B|$ fixed for the magnetic case.
\subsubsection*{Comments about Carrollian limits of general relativity} 

In both the scalar field and the Maxwell case we have seen that there are two types of Carroll limits. It is thus natural to wonder whether the same is true for general relativity. In \cite{Henneaux:1979vn} a Carroll limit of general relativity (in ADM variables) has been worked out leading to a theory with just kinetic terms (extrinsic curvature squared terms) as well as a cosmological constant, but thus without the spatial Ricci scalar term. The natural suggestion is that the other Carroll limit of general relativity requires a Lagrange multiplier term that sets the extrinsic curvature to zero on shell and which does contain a spatial Ricci scalar as well as a cosmological constant. We will report on this and other curved Carroll spacetime topics in \cite{dBHOSV3}.

\section{Perfect fluids \label{sec:PF} }
In the remainder of this paper we focus on applications of Carroll symmetry to cosmology and dark energy. We take a closer look at the arguments of the introduction, using the language of perfect fluids in a cosmological setting. Hence we will couple to gravity, in particular to the FLRW metric. 

We start with some general remarks about perfect fluids as discussed in \cite{deBoer:2017ing}. For perfect fluids with translation and rotation symmetry, but not necessarily boost symmetry, the energy-momentum tensor can be written as 
\begin{equation}\label{PFEMT}
T^t{}_t=-{\cal E}\ ,\qquad T^i{}_t=-({\cal E}+P)v^i\ ,\qquad T^t{}_j={\cal P}_j\ ,\qquad T^i{}_j=P\delta^i{_j}+v^i{\cal P}_j\ .
\end{equation}

This looks like the standard form of the energy-momentum tensor in a relativistic theory written in lab-frame coordinates, but it is more general, and holds also in the absence of any (i.e. Lorentz, Galilei or Carroll) boost symmetry. Here ${\cal E}$ and $P$ are energy density and pressure respectively, while  ${\cal P}_i$ is the momentum flow. Due to rotation symmetry, this quantity can be written as ${\cal P}_i=\rho v_i$ where $\rho$ is the kinetic mass density introduced in \cite{deBoer:2017ing}.


Though more general, the expression for the energy-momentum tensor in \eqref{PFEMT} can still be used for relativistic systems such as the real scalar field discussed in the introduction. In this case the energy-momentum tensor is given by \eqref{scalarstresstensor} in section \eqref{ZEL}
and we evaluate in the FLRW metric $g_{\mu\nu}={\rm diag}(-c^2, a^2(t) \delta_{ij})$. It is a straightforward exercise to put it in the form \eqref{PFEMT}, so that we read off
\begin{equation}
{\cal E}=\frac{1}{2}c^2\pi_\phi^2+\frac{1}{2a^2}|\nabla\phi|^2+V\ ,\qquad P=\frac{1}{2}c^2\pi_\phi^2-\frac{1}{2a^2}|\nabla\phi|^2-V\ ,\qquad \rho=\pi^2_\phi\ ,
\end{equation}
together with
\begin{equation}
v^i=-\frac{1}{\pi_\phi a^2}\partial_i\phi\ ,\qquad v_i=-\frac{1}{\pi_\phi}\partial_i\phi\ .
\end{equation}
Here $i,j$ indices are lowered or raised with the metric $h_{ij}=a^2\delta_{ij}$ or its inverse. Notice that ${\cal E}$ corresponds to the Hamiltonian density and the pressure $P$ to the Lagrangian in curved spacetime. The internal energy density is
\begin{equation}
\tilde{\cal E}\equiv {\cal E}-\rho v^2=-\frac{1}{2}\partial_\mu\phi\partial^\mu\phi+V(\phi)\ .
\end{equation}
It is Lorentz invariant and appears in the usual 
formulation of a relativistic perfect fluid tensor 
\begin{equation}
\label{TPFrel} 
T^\mu{}_\nu=\frac{\tilde{\cal E}+P}{c^2}U^\mu U_\nu+P\delta^\mu{_\nu}\ , 
\end{equation}
with relativistic four-velocities satisfying $U^\mu U_\mu=-c^2$. In fact, $\tilde{\cal E}$ is often denoted by $\rho$ in the literature, but we have reserved the symbol $\rho$ for the kinetic mass density. 

With $U^\mu=\gamma(1,\vec v)$ one can equate \eqref{TPFrel} to \eqref{PFEMT} and one can derive the general identities
\begin{equation}\label{E+P}
{\cal E}+P=c^2\rho\ ,\qquad {\cal E}+P=\gamma^2 ({\tilde{\cal E}}+P)\ ,
\end{equation}
where $\gamma^{-2}=1-v^2/c^2$ and $v^2=v^iv^jh_{ij}$. In a rest frame, in which $v^i=0$, we have that ${\cal E}=\tilde{\cal E}$ but this will not be used here. We note that the equations in \eqref{E+P}
were also shown in \cite{deBoer:2017ing} in flat spacetime, but as shown here they hold more generally. 
The first equation in \eqref{E+P} implies that in the limit of vanishing speed of light, the sum 
${\cal E}+P$ vanishes if the kinetic mass density remains finite. In the example of the scalar field, we have $\rho=\pi_\phi^2$ (for all velocities $v^i$) which is kept finite if we take the limit as in section \ref{ZEL}, so no rescaling of the scalar field. It thus follows that the energy density ${\cal E}=-P$ as a consequence of the limit $c\to 0$. The energy-momentum tensor in the Carroll limit and in the rest frame is then simply $T^\mu{}_\nu=P\delta^\mu{_\nu}$.

We can repeat this argument (for flat space) in more general terms based on \eqref{LorentzWard} and \eqref{CarrollWard}. Combined with \eqref{PFEMT}, we immediately find
\begin{equation}
 \frac{1}{c}T^i{}_t+cT^t{}_i=0\quad\rightarrow\quad{\cal E}+P=c^2\rho\ ,
\end{equation}
for Lorentz symmetry, whereas from imposing Carrollian symmetry we get 
\begin{equation}
 T^i{}_t=0\quad\rightarrow\quad{\cal E}+P=0\ ,
\end{equation}
provided that $v^i\neq 0$.
Notice that the Carroll case follows from the Lorentz case by taking $c\to 0$ while keeping $T^t{}_i$, hence the kinetic mass density $\rho$, finite, as before. Notice furthermore that this derivation only holds in a frame in which $v^i\neq 0$, but it can be shown that  ${\cal E}+P=0$ follows in any frame, i.e. also in the rest frame where $\vec v=0$. For this we use the covariant transformation law on the energy-momentum tensor,
\eqref{eq:trafolaws}. Substituting the energy-momentum tensor components \eqref{PFEMT} into this transformation law, it is an easy exercise to derive
\begin{equation}\label{boostrho}
{\cal E}'={\cal E}-({\cal E}+P)\vec b\cdot\vec v\ ,\quad P'=P\ ,\quad {\cal P}'_i=\Big({\cal P}_i-({\cal E}+P)b_i\Big)(1-\vec b\cdot\vec v)\ , \quad v'^{\,i}=\frac{v^i}{1-\vec b\cdot \vec v}\ .
\end{equation}
We know that the zero energy flux condition tells us that $({\cal E}+P)v^i=0$ so that this simplifies to 
\begin{equation}\label{boostrho2}
{\cal E}'={\cal E}\ ,\quad P'=P\ ,\quad {\cal P}'_i={\cal P}_i(1-\vec b\cdot\vec v)-({\cal E}+P)b_i\ , \quad v'^{\,i}=\frac{v^i}{1-\vec b\cdot \vec v}\ .
\end{equation}
Note that, as expected, the transformation of the velocity vector coincides with the one obtained in
 \eqref{transfvel}. 
 
Using the transformation law of the momentum density ${\cal P}_i$ and using $\mathcal{P}_i=\rho v^i$ we have straightaway for any $v^i$
\begin{equation}
    {\cal P}'_i=\rho'v'^i=\rho' \frac{v^i}{1-\vec b\cdot\vec v}=\rho v^i(1-\vec b\cdot\vec v)-b_i({\cal E}+P)\,,
\end{equation}
where we used the transformation law of the velocity. Since $v^i$ and $b_i$ are independent, solving this equation requires 
\begin{equation}
{\cal E}+P=0\,.
\end{equation}
While the vanishing of the energy flux only told us that the product of ${\cal E}+P$ and $v^i$ had to vanish we now see that it is always 
${\cal E}+P$ that must be zero, even when the velocity is zero. We emphasize that an essential ingredient here is the fact that the momentum density is of the form $\rho v^i$. As a byproduct of this calculation we also obtain the transformation of the kinetic mass density for cases with nonzero velocity, namely 
\begin{equation}
\rho'=\rho\, (1-\vec b\cdot \vec v)^2\ .
\end{equation}
We thus see that for any fluid velocity the equation of state of a Carroll perfect fluid is given by ${\cal E}+P=0$ in any frame, as announced in the introduction. 


The above argument relied heavily on the assumption that the momentum density is proportional to the fluid velocity, i.e. $\mathcal{P}_i=\rho v^i$. Ordinarily, a perfect fluid is described by a local temperature and a velocity field and this, together with rotational symmetry, is the origin of this assumption. However as we have seen for certain Carroll particles the momentum need not be related to the velocity, and so setting $v^i=0$ in \eqref{PFEMT} need not lead to a diagonal tensor but can instead give rise to an energy-momentum tensor that is of the form 
\begin{equation}
T^t{}_t=-{\cal E}\ ,\qquad T^i{}_t=0\ ,\qquad T^t{}_j={\cal P}_j\ ,\qquad T^i{}_j=P\delta^i{_j}\,,
\end{equation}
where the momentum density ${\cal P}_i$ is an independent variable. This energy-momentum tensor transforms as a Carroll tensor, i.e. as \eqref{eq:trafolaws} with $T^i{}_t=0$ provided ${\cal E}$ and $P$ are invariant under Carroll boosts and ${\cal P}_j$ transforms as ${\cal P}'_j={\cal P}_j-b_j\left({\cal E}+P\right)$. In this case the above argument no longer applies so that it no longer follows that ${\cal E}+P=0$. However, it is no longer clear that one should view this case as describing a fluid as it is not clear what the thermodynamic interpretation is. For more details we refer the reader to \cite{dBHOSV3} where we study Carroll fluids (on curved Carrollian spacetimes) in more detail.


\section{Friedmann equations and dark energy}\label{FE}

In the previous section we have shown that in the Carroll limit of a perfect fluid, one recovers an equation of state with ${\cal E}+P=0$, so $w=-1$. One may therefore expect that this leads to dark energy and exponential expansion of the universe when coupled to gravity. In this section, we show how this works by studying the Carroll limit of the Friedmann equations. As we will explain, some subtleties arise in this limit with regard to Newton's constant that needs to be rescaled properly before taking the limit $c\to 0$. Another subtlety that can arise is that in the Carroll limit, ${\cal E}+P=0$, but both ${\cal E}=P=0$. At the end of this section we illustrate how this can happen in a particular example starting from a relativistic fluid with $w=1$.

We start by coupling a relativistic perfect fluid to dynamical gravity, in our case the FLRW metric ${\rm d}s^2=-c^2{\rm d}t^2+a^2(t){\rm d}\vec x^{\,2}$. As is well known, this metric is written in comoving coordinates and so we work in the rest frame of the fluid where $v^i=0$, hence $\tilde{\cal E}={\cal E}$.  The Friedmann equations for a spatially flat universe in the rest frame of the fluid are then
\begin{equation}\label{FReqns}
\frac{\dot{a}^2}{a^2}=\frac{8\pi G_N}{3c^2}{\cal E}\ ,\qquad \frac{\ddot{a}}{a}=-\frac{4\pi G_N}{3c^2}\Big({\cal E}+3P\Big)\ ,
\end{equation}
where a possible cosmological constant has been absorbed in the pressure and energy density. Recall that the scale factor $a(t)$ is dimensionless. Using an equation of state of the form $P=w{\cal E}$ it follows that
\begin{equation}
\frac{\ddot{a}}{a}+\frac{1}{2}\frac{\dot{a}^2}{a^2}(1+3w)=0\ .
\end{equation}
Note that any explicit dependence on  $c$ has dropped out, so the Carroll limit is trivial here. The solutions are well known and one must separate $w=-1$ from the rest: 
\begin{eqnarray}\label{solscale}
&&w\neq -1:\qquad a(t)=(a_0t+a_1)^{\frac{2}{3(1+w)}}\ ,\qquad H(t)\equiv \frac{\dot a}{a}=\frac{2}{3(1+w)}\frac{a_0}{a_0t+a_1}\nonumber\\
&&w=-1: \qquad a(t)=a_0e^{Ht}\ ,
\end{eqnarray}
with two integration constants $a_0$ and $a_1$, and $H(t)$ the Hubble function which is constant for $w=-1$. 
The Hubble radius, $R_H(t)=cH^{-1}$ grows linear in time for any value of $w\neq -1$.

From the Friedmann equations follow the identities, valid for any $w$,
\begin{equation}\label{FE2}
{\cal E}+P=\frac{3c^2}{8\pi G_N} (1+w)H^2(t)\qquad \rightarrow \qquad \rho=\frac{3}{8\pi G_N} (1+w)H^2(t)\ ,
\end{equation}
where we remind that the kinetic mass density $\rho$ follows from \eqref{E+P}. One can see from the first equation that ${\cal E}+P\to 0$ in the Carroll limit, but some care is needed with specifying what is kept fixed in the limit $c\to 0$. To illustrate this, we look at two examples, one with $w=-1$, and one with $w=+1$. We focus again on the case of a single scalar field, for which we have $\rho=\pi^2_\phi=(\partial_t \phi)^2/c^4$. 

The simplest case is dark energy, $w=-1$, i.e. a scalar field constant in time and space, so $\pi_\phi=0$ and $v_i=0$, but with a nonzero, but constant potential $V=\Lambda={\cal E}$, such that the Klein-Gordon equation is satisfied.  The metric is that of de Sitter spacetime with a horizon given by the Hubble radius. We then find, from the first equation in \eqref{FReqns},
\begin{equation}\label{H0}
H^2=\frac{8\pi G_N}{3c^2}\Lambda\ .
\end{equation}
For fixed $\Lambda$ and $G_N$, the Hubble constant would diverge in the Carroll limit, but it is important that $H$ stays fixed in order to maintain exponential expansion and the conformal transformations as isometry group (see \eqref{ConfCarroll}). Furthermore, as explained in the introduction we want to keep the potential $V=\Lambda$ fixed and finite in the Carroll limit and so we need to rescale Newton's constant such that in the Carroll limit, the Hubble constant is kept fixed,
\begin{equation}\label{eq:gnc}
G_C\equiv \frac{G_N}{c^2}\ , \qquad H^2=\frac{8\pi G_C}{3}\Lambda\ ,\qquad G_C\,\,\,\rm{fixed}\ .
\end{equation}
The Hubble radius $R_H$ then goes to zero, $R_H=cH^{-1}=\frac{c}{{\sqrt{8\pi G_C\Lambda}}}\to 0$, as desired in the Carroll limit\footnote{By vanishing Hubble radius, we mean it is smaller than any other length scale in the problem. In empty de Sitter, there is however no other length scale, so we rescale $R_H\to \epsilon R_H$ and send $\epsilon\to 0$. We remark furthermore that our analysis is classical. In a quantum theory, one could compare with the Planck length. Then the classical Carroll regime is valid for small Hubble radius, but still larger than the Planck length.}.

 We have ${\cal E}+P=0$ in the Carroll limit (because $w=-1$), with both ${\cal E}$ and $P$ finite and nonzero, i.e. ${\cal E}=\Lambda$. So we can confirm the picture raised in the introduction that the Carroll limit builds up the spacetime from small Hubble cells that grow with growing values of $c$. The Hawking temperature is constant throughout de Sitter space, and should stay fixed in the Carroll limit where we rescale the size of the Hubble patch. The entropy however is expected to vanish, since it is associated to the area of the horizon which vanishes in the Carroll limit. Indeed, inserting all SI-units, we have
\begin{equation}\label{eq:gibbonshawking}
k_BT=\frac{H}{2\pi}\hbar\ ,\qquad S_{GH}=\frac{k_Bc^3}{4\hbar G_N}A=\frac{\pi k_B}{\hbar G_C}\frac{c^3}{H^2}\to 0\ .
\end{equation}
Here we have assumed that $k_B$ and $\hbar$ behave the same in the Carroll limit.
The conclusion of this is that de Sitter space in the Carroll limit becomes conformal to ${\mathbb R}^3$,  with metric ${\rm d}s^2=e^{Ht}{\rm d}{\vec x}^{\,2}$. Expanding around $c=0$ opens up Hubble patches  with radius $cH^{-1}$, within which the temperature is constant, but the entropy scales with $c^3$.

Next we consider the second example, with $w\neq -1$, say  $w=1$, a free scalar field with vanishing potential $V=0$. What happens when we take the zero speed of light limit? We cannot expect dark energy or inflation, yet there should be a well-defined Carroll limit with ${\cal E}+P=0$, suggesting $w=-1$. It is actually not difficult to figure out the solution to this apparent paradox. The Klein-Gordon equation is $\ddot\phi+3H\dot\phi=0$ for zero potential, or in terms of momenta 
\begin{equation}
\dot\pi_\phi+3H\pi_\phi=0\ .
\end{equation}
This is easily solved by 
\begin{equation}
\pi_\phi=\sqrt{\frac{1}{12\pi G_N}}\frac{1}{(t+a_1/a_0)}\ ,
\end{equation}
where we took into account \eqref{FE2}, i.e. $\rho=\pi_\phi^2=3H^2/(4\pi G_N)$, as well. The solution for the scalar field is then
\begin{equation}
\phi(t)=\frac{c^2}{\sqrt{12\pi G_N}}\ln \Big( t +a_1/a_0\Big) +\phi_0\ .
\end{equation}
The SI-units for $\phi$ (in 3+1 dimensions) are ${(kg\cdot m)}^{1/2}s^{-1}=J^{1/2}m^{-1/2}$. In the Carroll limit, there is no 
need to rescale $G_N$ with a factor $c^2$ this time, as $H$ is already finite in the limit (see \eqref{solscale}). Therefore, $\phi$ goes to a constant in the $c\to 0$ limit, but the momentum does not, it goes like $t^{-1}$. The energy density is 
\begin{equation}
{\cal E}= \frac{1}{2}c^2\pi_\phi^2\  \to 0\ .
\end{equation}
It goes like ${\cal E}\sim c^2 t^{-2}$ and vanishes in the Carroll limit. Similarly the pressure vanishes in the Carroll limit. So what we find is that ${\cal E}+P$ vanishes without $w=-1$, (in fact $w=+1$) but the reason is that both ${\cal E}$ and $P$ vanish. We therefore see that the equation ${\cal E}+P=0$ does not always mean dark energy (exponential growth), because both energy and pressure can be  zero. But note that even for  vanishing energy and pressure the scale factor is nontrivial and still the same as for $c \neq 0$; this is because the $c^2$ dependence in ${\cal E}$ and $P$ cancels against the $c^{2}$ factor in the denominator of \eqref{FReqns}. The reason why such an evolution for the scale factor can take place is alluded to in the introduction: even for non-inflationary metrics, there are (time-dependent) super-Hubble scales and superluminal recession velocities. A non-trivial Carroll limit should therefore still exist.

\section{Inflation}\label{Inflation}

In the previous section, we looked at two examples with a constant value of $w$. Generically, however, $w$ is time dependent, as is the case during inflation. In this section, we address what happens to inflation in the Carroll limit. As we will show, the Carroll limit enforces the limit $w \to -1$, so towards a de Sitter phase. We will analyze in detail the Carroll limit of chaotic inflation (with a quadratic potential), but we expect our conclusions to be more general.

Before we consider the limit, let us quickly recap the chaotic inflation scenario. Ignoring spatial derivatives, which we address in the next subsection, the scalar field equation of motion is $\ddot\phi+3H\dot\phi=-c^2\frac{\partial V}{\partial \phi}$. For only a mass term, the potential is $V=\frac{1}{2}\frac{m^2c^2}{\hbar^2}\phi^2$ in SI-units. The first of the Friedmann equations in \eqref{FReqns} can now be written as
\begin{equation}
\label{Heq} 
H^2=\frac{4\pi G_N}{3}\Big(\pi^2_\phi+\frac{m^2\phi^2}{\hbar^{2}}\Big)
\ ,
\end{equation}
with $\pi_\phi=\frac{1}{c^2}\dot\phi$.
The scalar field equation is
\begin{equation}\label{KGInflation}
\dot\pi_\phi+3H\pi_\phi+\frac{m^2c^2}{\hbar^2}\phi=0\ .
\end{equation}
In chaotic inflation, we start with an initial condition for which the scalar field is very large. This means that $H$ should be very large as well. One furthermore looks for solutions in which $\pi_\phi$ is small at early times compared to the $\phi$ and $H$ terms  in \eqref{Heq} which are large at early times. So $\phi$ varies slowly in time. We also assume that $\dot \pi_\phi$ is small at early times and can be ignored in \eqref{KGInflation}. This then leads to an approximate solution at early times
\begin{equation}\label{solchaotinfl}
\phi=\phi_0-\frac{c^2}{ {\sqrt{12\pi G_N}}}\,\frac{mc^2}{\hbar}\,t\ ,\qquad  H=
{\sqrt{\frac{4\pi G_N}{3}}}\frac{m}{\hbar}\phi\ ,
\end{equation}
which is in the textbooks on chaotic inflation. The second Friedmann equation is also satisfied, it can be written as $H^2+\dot H=\frac{4\pi G_N}{3}\Big(\frac{m^2\phi^2}{\hbar^2}-2\pi_\phi^2\Big)$, where we can ignore $\dot H$ and $\pi_\phi$ at early times. So one can see that at early times, one starts off in a de Sitter phase (constant $H$), and the linear terms in $t$ deviate away from this. One can now check that dropping the $\pi_\phi^2$-term in \eqref{Heq} is justified when the Hubble constant $H_0=H(t=0)$ is much larger than the scalar mass $m$ in appropriate units. 

Now we reconsider the equations in the Carroll limit and in the spirit of section \ref{ZEL} we make expansions\footnote{One could, as in section \ref{ZEL}, introduce additional overall scaling factors $c^{\Delta_\phi}$ and $c^{\Delta_H}$ in front of the expansion, but we did not find any other interesting, inequivalent, solutions. Moreover, we want that de Sitter solution should be included, where $H$ is constant and nonzero in the $c\to 0$ limit.}
\begin{equation}\label{expHubble}
    \phi=\phi_0+c^2\phi_1+\cdots\ ,\qquad \pi_\phi=\frac{1}{c^2}\dot\phi_0+\dot\phi_1+\cdots\ ,\qquad H=H_0+c^2H_1+\cdots\ ,
\end{equation}
with all coefficients time dependent. Furthermore, we keep fixed the combinations
\begin{equation}
    G_C\equiv \frac{G_N}{c^2}\ ,\qquad \mu\equiv \frac{mc}{\hbar}\ ,
\end{equation}
in the limit $c\to 0$ such that the potential $V=\frac{1}{2}\mu^2\phi^2$ stays finite and we can have nontrivial solutions. The leading order terms in the field equations then start at $c^{-2}$ and require $\dot\phi_0=0$, and at order $c^0$, we get
\begin{equation}\label{H0phi0}
    H_0^2=\frac{4\pi G_C}{3}\mu^2\phi_0^2\ ,\qquad \ddot \phi_1+3H_0\dot\phi_1=-\mu^2\phi_0\ .
\end{equation}
The first equation determines the leading order solution for the Hubble factor and the scalar field, and corresponds to a de Sitter solution (since $\phi_0$ is constant in time). The second equation perturbs the scalar field away from being constant. The solution  of the inhomogeneous equation is given by
\begin{equation}
    \phi_1(t)=-\frac{\mu^2}{3H_0}\phi_0t\ .
\end{equation}
Possible solutions to the homogeneous equation are not included, as they can be absorbed in the constant $\phi_0$ or set to zero by appropriate boundary conditions. With these boundary conditions, one reproduces precisely the inflationary solution for the scalar field in \eqref{solchaotinfl}. At order $c^2$ in the Friedmann equation, one determines $H_1$ and we find
\begin{equation}
    H_1=\frac{\mu^2}{18H_0}-\frac{1}{3}\mu^2t\ ,
\end{equation}
which leads to
\begin{equation}
    H=H_0\Big(1+\frac{c^2\mu^2}{18H_0^2}\Big)-\frac{1}{3}\mu^2c^2t\ .
\end{equation}
Notice that this only matches the result from inflation (the second equation in \eqref{solchaotinfl}) when $\mu c \ll H_0$. This condition was needed for the validity of the approximation made in inflation, but is not needed in the derivation of the Carroll expansion. Of course we can redefine $H_0$ but the relation with $\phi_0$ as given by \eqref{H0phi0} is then lost.

The momentum $\pi_\phi$ stays constant in the Carroll limit, $\pi_\phi=-\frac{\mu^2}{3H_0}\phi_0$ and the energy density goes like
\begin{equation}
{\cal E}=\frac{1}{2}c^2\pi_\phi^2+V(\phi) \rightarrow \frac{1}{2}\mu^2\phi_0^2\ ,
\end{equation}
and $P=-{\cal E}$. So we conclude that in the Carroll limit, inflationary solutions are attracted to dark energy solutions with an equation of state $w=-1$.

The above analysis can be translated in terms of the slow roll parameters. For chaotic inflation, $\epsilon=\eta$, with, in SI units,
\begin{equation}
    \epsilon=\frac{1}{2}\frac{c^4}{G_N}\Big(\frac{V'}{V}\Big)^2=\frac{c^4}{G_N}\frac{2}{\phi^2}\ .
\end{equation}
Using \eqref{solchaotinfl}, we can rewrite the slow roll parameter as
\begin{equation}
    \epsilon=\frac{8\pi}{3}\frac{R_H^2(t)}{\lambda^2}\ ,
\end{equation}
so that slow roll is guaranteed for all times for which the Compton wavelength is larger than the Hubble radius, $\lambda \gg R_H(t)$, so for super-Hubble scales. The Carroll limit guarantees this because $R_H\to 0$ for $c\to 0$.

\section{Scalar perturbations in de Sitter spacetime}\label{sec:PS}

We now look at perturbations of the fields, and in particular focus on scalar perturbations. In this subsection, we focus on the massless case. This is standard analysis and we follow the notation of \cite{Baumann:2009ds}. The only minor difference in the notation is that we reintroduce the speed of light $c$ into the expressions. An important set of quantities in the perturbation analysis are the Bunch-Davies mode functions $v_k$ at given wavenumber $k$ appearing in the plane wave expansion of the scalar perturbation around a de Sitter background. Plane waves are given by $e^{i\vec k\cdot \vec x}$, so $k$ has inverse length. The solutions for these mode functions in a de Sitter background are given by (see e.g. eq. (196) in \cite{Baumann:2009ds}),
\begin{equation}\label{modefunctions}
    v_k=\frac{e^{-ikc\tau}}{{\sqrt{2k}}}\Big(1-\frac{i}{kc\tau}\Big)\ .
\end{equation}
Here $\tau$ here is conformal time, which is given by $\tau=-1/(aH)$ in de Sitter. In the far past, $\tau \to -\infty$, modes are supposed to start to be sub-Hubble with $|kc\tau|\gg 1 $, which for de Sitter means $k^{-1}\ll c/(aH)\equiv {\cal R}_H$, the comoving Hubble radius. So for early-time-modes within the comoving Hubble sphere, the modes scale like $k^{-1/2}$ and oscillate like in a Minkowski spacetime. The comoving Hubble radius however shrinks in time, as
\begin{equation}
    {\cal R}_H=\frac{c}{H}e^{-Ht}\ ,
\end{equation}
so the mode will at some point later in time exit the horizon and become super-Hubble, $k^{-1}> {\cal R}_H$, or in terms of the wavelength $\lambda\equiv a/k>R_H$, or equivalently, $|kc\tau|<1$. In  this regime, the second term takes over and the mode functions will scale like $k^{-3/2}$. For the rescaled fields $\psi_k\equiv \frac{v_k}{a}$, this implies that they freeze in time at super-horizon scales. 

The important observation here is that the exiting of the mode functions from the horizon is also achieved in the Carroll limit $c\to 0$, as expected since this limit would also shrink the comoving Hubble radius. Indeed, in the Carroll limit, the second term in \eqref{modefunctions} dominates, and we get straightaway
\begin{equation}
    \psi_k\equiv \frac{v_k}{a}\stackrel{c\to 0}{\rightarrow} \frac{i}{\sqrt 2}\frac{H}{k^{3/2}}\ ,
\end{equation}
which now is time-independent. 
So in order words, the Carroll limit gives us the behaviour of the mode-functions after crossing time. Therefore, also the correlation functions freeze out at super-Horizon scales.

\subsection{Massive scalar in de Sitter}

We now consider a massive scalar field in de Sitter and switch on the spatial derivatives as perturbations in the scalar field. After Fourier transforming the spatial components, the Klein-Gordon field equation becomes 
\begin{equation}
    \ddot \phi_k+3H\dot\phi_k+\Big(\omega_0^2+\frac{c^2k^2}{a^2}\Big)\phi_k=0\ ,
\end{equation}
with wave-vector as in $\phi\sim \phi_k(t) e^{i\vec k\cdot \vec x}$ and as before, $\omega_0\equiv \mu c=mc^2/\hbar$ and $k^2=k^i\delta_{ij}k^j$. 
Notice that the perturbation in spatial derivatives comes with a factor of $c^2$, so it is suppressed to leading order in the Carroll expansion. Furthermore, we now keep $\omega_0$ fixed in the Carroll limit. The potential (mass term) itself would then vanish in the strict Carroll limit, but this is allowed as this massive scalar field need not be the inflaton field, and the de Sitter background is already fixed, so we take $H$ constant. 

As before, we make an expansion around small values of $c$, 
\begin{equation*}
    \phi=\phi_0+c^2\phi_1+\cdots\ ,
\end{equation*}
and find to leading order
\begin{equation}
     \ddot \phi_{k,0}+3H\dot\phi_{k,0}+\omega_0^2\phi_{k,0}=0\ ,
\end{equation}
which is easily solved by
\begin{equation}
    \phi_{k,0}=f_{\pm} e^{\lambda_\pm Ht}=f_\pm a^{\lambda_\pm}\ ,\qquad \lambda_\pm=-\frac{3}{2}\pm{\sqrt{\frac{9}{4}-\frac{\omega_0^2}{H^2}}}\equiv -\frac{3}{2}\pm \nu_\phi\ .
\end{equation}
The functions $f_\pm $ can depend on $k$ and can be determined by normalization conditions, similar to the massless case. Observe however that now there is a small time-dependence, and the fluctuations do not freeze out in the strict Carroll limit. In the zero mass limit, they however do, as $\omega_0\to 0$ and $\lambda_+\to 0$, consistent with the massless case if we choose boundary conditions that dissallow the $\lambda_-$ solution (which goes like $e^{-3Ht}$ in the massless case and diverges when $t\to -\infty$).

At next order, the equation for $\phi_1$ is 
\begin{equation}
     \ddot \phi_{k,1}+3H\dot\phi_{k,1}+\omega_0^2\phi_{k,1}=-\frac{k^2}{a^2}\phi_{k,0}=-k^2f_+(k)e^{(\lambda_+-2) Ht}\ ,
\end{equation}
which is solved by (dropping again integration constants)
\begin{equation}
    \phi_{k,1}=-\frac{k^2}{2a^2H^2}\frac{\phi_{k,0}}
    {2\Big(1-\nu_\phi\Big)}\ .
\end{equation}
For $\nu_\phi=1$, this solution does not hold and is replaced by
\begin{equation}
    \phi_{k,1}=\frac{k^2f_+}{4H^2}(1+2Ht)e^{-\frac{5}{2}Ht}\ .
\end{equation}
Grouping things together, we find for the solution for the massive scalar in the Carroll expansion (with $\nu_\phi\neq 1$)
\begin{equation}
    \phi_k=f_+(k)\,a^{\lambda_+}\Big(1-\frac{c^2k^2}{4a^2H^2}\frac{1}{1-\nu_\phi}+\cdots\Big)\ .
\end{equation}
The leading term proportional to $a^{\lambda_+}=\Big(\frac{1}{a}\Big)^{\frac{3}{2}-\nu_\phi}$ is well-known from the literature and appears in the power spectrum as well.

\section{Outlook}\label{sec:outlook}

In this paper we have explored various aspects of Carroll symmetry. The Carroll symmetry
algebra arises in the $c\rightarrow 0$ limit of the Poincar\'e algebra. As it stands, this
limit appears to be very unphysical, as particles with finite velocities become tachyonic, the
usual Lorentz factor $\sqrt{1-v^2/c^2}$ becomes imaginary, etc. While it is true that physical
velocities can never be relevant for the $v \gg c$ regime, there can be other effective 
velocities in a system
for which this condition holds, and for which the Carroll limit and the small $c$ expansion
are physically meaningful. The prime example that we considered in this paper is that of
superluminal recessional velocities in expanding universes. We showed that this is 
relevant for inflation and argued even for non-inflationary models. This led us
to advocate that this perspective offers the potential of understanding the universe as an
expansion in $c$ around a Carroll point with Carroll symmetry. As an example, we saw that
the de Sitter group of isometries in the limit becomes the conformal group in Euclidean 
flat space, consistent with a holographic interpretation of dark energy, and a key ingredient
in the bootstrap approach to cosmological correlators \cite{Maldacena:2011nz,Arkani-Hamed:2018kmz}.

A second situation in which the $c\rightarrow 0$ limit is meaningful is when the relevant
``effective velocity" is related to the time dependence of classical field configurations. 
For these there is no issue organizing the theory as an expansion in $c$ as we illustrated in
various examples in section~\ref{sec:CFT}.

An alternative perspective on the role of the $c\rightarrow 0$ limit is that in this limit lightcones
get squeezed and therefore spatially separated points become causally disconnected. In the 
cosmological context it is the different Hubble patches which become causally disconnected. This
suggests a possible broader applicability of the Carroll limit to other situations where points
are taken to be approximately causally disconnected, in particular to all situations where 
one attempts to define an S-matrix for localized asymptotic states. Such localized 
asymptotic states are assumed to exist in isolation and not interact with each other at early 
and late times, and may only exists approximately in actual quantum field theories, for 
example due to IR divergences.
In such cases one can try to factorize the S-matrix in a hard and soft part and we speculate that 
Carroll-like limits may in general be of relevance to the hard part of such amplitudes, but that the $E=0$ Carroll particles could be of relevance to the soft modes.
We leave a further investigation of this issue to future work.

The problems that arise in taking the $c\rightarrow 0$ limit at the level of individual point
particles also show up when trying to interpret energy-momentum tensors that are compatible with the
Carroll symmetries as describing the thermodynamics of a well-defined underlying quantum system. 
The Carroll algebra does not allow one to write a non-trivial dispersion relation which relates
the energy to the momentum, and this leads to a divergence in the integral over momenta in the
finite temperature partition function\footnote{This same conclusion would apply to massless Galilean particles as well, since in that case only the Hamiltonian would transform under Galilean boost, and not the momentum. It is in fact the mass, or availability of a central extension, in the Bargmann case which enables momentum to transform as well under Galilean boost. In general dimensions, a central extension like this is unavailable for Carroll, see e.g. \cite{Bergshoeff:2014jla}.}. One can try to 
avoid this conclusion in several ways, e.g. by restricting the integral over momenta by hand,
or looking at systems with no momenta such as spacetime-filling branes, but none of these lead
to a particularly compelling picture. One could also choose to take the Carroll limit directly
at the level of the partition function of a relativistic gas, but this leads to a vanishing result
unless one adds additional factors of $c$ by hand. In such a case one gets a finite answer which
defies a direct quantum mechanical understanding unless one is willing to consider e.g. a gas
of tachyons, but we do not think that this is a particularly interesting direction to explore for
obvious reasons. We refer the reader to our forthcoming work \cite{dBHOSV3} in which we elaborate
on several of these points. 

Another observation that we would like to highlight are the two types of Carroll limits which
generically seem to exist. These two different limits are already visible in the Carroll algebra
where the representation theory is quite different depending on whether the energy vanishes or not.
These two qualitatively different behaviours also appear if we look at correlation functions. 
Consider for example the two-point function of two Carroll scalar fields. It is easy to see that 
the following two answers are both solutions to the Carroll Ward identities
\be
\langle O(t,\vec{x}) O(0,0) \rangle = f(|\vec{x}|)
,\qquad
\langle O(t,\vec{x}) O(0,0) \rangle = F(t)\delta(\vec{x})
\ee
where the first case corresponds to vanishing energy, and the second one to non-vanishing energy.
Equivalently, the first case is one where we put the canonical momentum of a field equal to zero,
whereas the second case is relevant for theories where we drop spatial derivatives, in line with
the field theory limits considered in section~\ref{sec:CFT}.

To quote Lewis Carroll from Alice in Wonderland: “Begin at the beginning," the King said, very gravely, “and go on till you come to the end: then stop.” We tried to follow the advice 
of the King quite closely in this paper and hope to have convinced the reader that we have
not quite come to the end (yet).

\subsection*{Acknowledgments}


We thank Eric Bergshoeff, Joaquim Gomis, Gerben Oling and Guilherme L. Pimentel for useful discussions.
JdB is supported by the European Research Council under the European Unions Seventh Framework Programme (FP7/2007-2013), ERC Grant agreement ADG 834878.
JH is supported by the Royal Society University Research Fellowship ``Non-Lorentzian Geometry in Holography'' (grant number UF160197). 
NO is supported in part by the project ``Towards a deeper understanding of black holes with non-relativistic holography'' of the Independent Research Fund Denmark (grant number DFF-6108-00340) and by the Villum Foundation Experiment project 00023086. 
WS is supported by the Icelandic Research Fund (IRF) via a Personal Postdoctoral Fellowship Grant (185371-051).

\appendix
\section{Further details on representations of the Carroll algebra}\label{app:irreps}

The Carroll algebra consists of the generators $H$ (Hamiltonian), $P_i$ (spatial momenta), $C_i$ (Carroll boosts) and $J_{ij}=-J_{ji}$ (spatial rotations). The nonzero commutators are given by
\begin{eqnarray}
&&[P_i\,, C_j]=\delta_{ij}H\,,\\
&& [J_{ij}\,, P_k] = \delta_{ik}P_j-\delta_{jk}P_i\,,\\
&& [J_{ij}\,, C_k] = \delta_{ik}C_j-\delta_{jk}C_i\,,\\
&& [J_{ij}\,, J_{kl}] = \delta_{ik}J_{jl}-\delta_{jk}J_{il}+\delta_{jl}J_{ik}-\delta_{il}J_{jk}\,,
\end{eqnarray}
where $i,j,k,l$ run over $1,\ldots, d$. Let us define $M_{ij}$ as
\begin{equation}
M_{ij}=HJ_{ij}+C_i P_j-C_j P_i \,.
\end{equation}
This quantity satisfies
\begin{eqnarray}
&& [M_{ij}\,,P_k]=0\,,\qquad [M_{ij}\,,C_k]=0\,,\\
&&[M_{ij}\,,J_{kl}]=\delta_{ik}M_{jl}-\delta_{jk}M_{il}+\delta_{jl}M_{ik}-\delta_{il}M_{jk}\,.
\end{eqnarray}
These results imply that $M^2=\frac{1}{2}M_{ij}M_{ij}$ is a Casimir of the Carroll algebra. We can write out $M^2$ to obtain
\begin{equation}\label{eq:Msquared}
M^2=\frac{1}{2}H^2J_{ij}J_{ij}+2HJ_{ij}C_iP_j+C_iC_iP_jP_j-(C_i P_i)^2-(d-2)HC_i P_i\,.
\end{equation}
A useful result is
\begin{equation}\label{eq:Mcommutator}
[M_{ij}\,, M_{kl}]=H\left(\delta_{ik}M_{jl}-\delta_{jk}M_{il}+\delta_{jl}M_{ik}-\delta_{il}M_{jk}\right)\,.
\end{equation}

For $d=3$ it is useful to define $W_k$ and $S_k$ via
\begin{equation}
M_{ij}=\varepsilon_{ijk}W_k\,,\qquad J_{ij}=\varepsilon_{ijk}S_k\,,
\end{equation}
where $\varepsilon_{ijk}$ is the Levi-Civita symbol. We then find the Carroll algebra
\begin{eqnarray}
[P_i\,, C_j]=\delta_{ij}H\,,\qquad [S_i\,, P_j]=\varepsilon_{ijk}P_k\,,\qquad [S_i\,, C_j]=\varepsilon_{ijk}C_k\,,\qquad [S_i\,, S_j]=\varepsilon_{ijk}S_k\,.
\end{eqnarray}
Furthermore, $W_i=HS_i-\varepsilon_{ijk}P_j C_k$ and it obeys the following commutators
\begin{equation}
[P_i, W_j]=0\,,\qquad [C_i\,,W_j]=0\,,\qquad [S_i, W_j]=\varepsilon_{ijk} W_k\,,\qquad [W_i\,, W_j]=\varepsilon_{ijk}HW_k\,.
\end{equation}
In $d=3$ we can also define the operator $L=S_iP_i$ (which is closely related to the helicity operator) and which obeys
\begin{equation}\label{eq:L}
[L\,, P_i]=0\,,\qquad [L\,, S_i]=0\,,\qquad [L\,,C_i]=W_i\,,\qquad [L\,, W_i]=-\varepsilon_{ijk}P_j W_k\,.
\end{equation}

The Carroll algebra is a semi-direct sum of the Abelian ideal spanned by $\{H, P_i\}$ and the Euclidean algebra $\mathfrak{iso}(d)$ spanned by $\{J_{ij}, C_i\}$. We will consider induced representations using the little group method. 

States in our representation space will be eigenstates of the central element $H$ and the Casimir $M^2$. Consider eigenstates whose energy $E$ is nonzero. Under a Carroll boost the momentum transforms as $\vec {p'}=\vec p-\vec b E$ where $\vec b$ is the Carroll boost transformation parameter. When $E\neq 0$ we can thus always transform to a frame in which $\vec p=0$. On such states, using that
\begin{equation}
M_{ij}=HJ_{ij}-P_i C_j+P_j C_i=HJ_{ij}-C_j P_i +C_i P_j \,,
\end{equation}
the $M_{ij}$ act as $EJ_{ij}$. In other words 
on such states $M_{ij}=EJ_{ij}$ form an $\mathfrak{so}(3)$ algebra as follows from \eqref{eq:Mcommutator}. The eigenvalues of the Casimir $M^2$, using \eqref{eq:Msquared}, are $E^2s(s+1)$ where $s(s+1)$ is the eigenvalue of $\frac{1}{2}J_{ij}J_{ij}=S_i S_i$ with $s=0,\frac{1}{2},1,\ldots$. The little group $SO(3)$ of these nonzero energy states is the invariance group of the energy-momentum eigenstates $(E\neq 0, \vec p=0)$.

Next we consider states with energy $E=0$ and we demand $\vec P\cdot \vec P>0$. For zero energy states, the momentum is a Carroll boost invariant object and so each energy-momentum eigenstate is of the form $(E=0, \vec p\neq 0)$. On such states the vector whose components are $ W_i=HS_i+\varepsilon_{ijk}C_j P_k$ is of the form $\vec C\times\vec P$. and they form a 2-dimensional Abelian algebra as $\vec P\cdot \vec W=0$. Rotations around the momentum axis also leave $(E=0, \vec p\neq 0)$ invariant and so the little group is the Euclidean group $ISO(2)$ given by $L$ and $W_i$ whose commutators are $[W_i\,, W_j]=0$ and $[L\,, W_i]=-\varepsilon_{ijk}P_j W_k$. The Casimir $M^2$ acts on such states as $(\vec C\cdot\vec C)(\vec P\cdot\vec P)-(\vec C\cdot\vec P)^2$.

There are now two cases. Case I: $\vec W\cdot \vec W=0$ so that $\vec W=0$. Case II: $\vec W\cdot\vec W>0$. Strictly speaking there is a third case with $\vec P\cdot \vec P=0$ for all zero energy states, but this case corresponds to the vacuum as all operators return zero for such states. We observe that for case I the eigenvalue of the helicity operator $\frac{1}{\sqrt{\vec P\cdot \vec P}}\vec S\cdot\vec P$ is also a good label as $\frac{1}{\sqrt{\vec P\cdot \vec P}}\vec S\cdot\vec P$ commutes with all operators of the algebra when acting on zero energy states. This follows from equation \eqref{eq:L}. The labels for the zero energy states corresponding to case I are the eigenvalues of $\vec P\cdot \vec P>0$, and the helicity operator.

\bibliographystyle{JHEP}
\bibliography{main}
\end{document}